\documentclass[paper,11pt]{article}
\pdfoutput=1

\usepackage{comment}
\usepackage{graphicx}
\usepackage[centertags]{amsmath}
\usepackage{amsfonts}
\usepackage{amssymb} 
\usepackage{amsthm}
\usepackage{slashed}
\usepackage{enumerate,enumitem}
\usepackage{amsbsy}
\usepackage{bbm}
\usepackage{mathrsfs}
\usepackage{url}
\usepackage{bbm}
\usepackage{color}
\usepackage{framed}
\usepackage{subfig}
\usepackage{caption}
\usepackage{float}
\usepackage{cancel}
\usepackage{epstopdf}
\usepackage{jheppub}
\usepackage{hyperref}

\def\({\left(}
\def\){\right)}
\def\[{\left[}
\def\]{\right]}

\newcommand{\ket}{\rangle}

\def\one{{\rm 1\kern -.9mm l}}                             %

\def\beq{\begin{equation}}
\def\eeq{\end{equation}}
\def\beqa{\begin{eqnarray}}
\def\eeqa{\end{eqnarray}}
\newcommand{\eqa}{\begin{eqnarray}}
\newcommand{\ena}{\end{eqnarray}}

\newcommand{\cL}{\mathcal{L}}

\newcommand{\cS}{\mathcal{S}}

\newcommand{\uZ}{u}

\DeclareMathAlphabet{\mathpzc}{OT1}{pzc}{m}{it}

\DeclareMathOperator{\sech}{sech}
\newcommand\blank[1]{}
\newcommand{\fract}[2]{{\textstyle\frac{#1}{#2}}}

\newcommand\RR{{\mathbb R}}

\newcommand{\balpha}{\alpha\kern -6.7pt\alpha}
\newcommand{\bbalpha}{\alpha\kern -4.95pt\alpha}

\newcommand\en{\end{equation}}
\newcommand\bea{\begin{eqnarray}}
\newcommand\eea{\end{eqnarray}}

\newcommand{\One}{{\hbox{{\rm 1{\hbox to 1.5pt{\hss\rm1}}}}}}
\renewcommand{\One}{{\mathbb 1}}
\renewcommand{\One}{{\rm 1\!\!1}}

\newcommand{\ba}{\begin{eqnarray}}
\newcommand{\ea}{\end{eqnarray}}

\newcommand{\be}{\begin{equation}}
\newcommand{\ee}{\end{equation}}

\renewcommand{\log}{\ln}

\newcommand{\TbT}{\text{T}\bar{\text{T}}}
\newcommand{\JbT}{\text{J}\bar{\text{T}}}

\newcommand{\ta}{\tau}

\renewcommand{\ell}{{\mathcal L}}

\setlist[itemize]{leftmargin=*}

\def\XXint#1#2#3{{\setbox0=\hbox{$#1{#2#3}{\int}$ }
\vcenter{\hbox{$#2#3$ }}\kern-.6\wd0}}

\title{\boldmath The $\TbT$ perturbation and its geometric interpretation }

\author[1]{Riccardo Conti,}
\author[2]{Stefano Negro,}
\author[1]{Roberto Tateo}

\affiliation[1]{Dipartimento\ di Fisica and Arnold-Regge Center, Universit\`a di Torino, and INFN, Sezione di Torino, Via P. Giuria 1, I-10125 Torino, Italy.}
\affiliation[2]{ C.N. Yang Institute for Theoretical Physics, New York Stony Brook, NY 11794-3840. U.S.A.}

\emailAdd{riccardo.conti@to.infn.it}
\emailAdd{stefano.negro@stonybrook.edu}
\emailAdd{tateo@to.infn.it}
\abstract{
Starting from the recently-discovered $\TbT$-perturbed Lagrangians, we prove
that  the  deformed   solutions to the classical EoMs for bosonic  field theories  are equivalent to the unperturbed ones but for  a specific field-dependent local change of coordinates. 
This surprising geometric outcome is fully consistent  with the identification of $\TbT$-deformed 2D quantum field theories as  topological JT gravity  coupled to generic matter fields. Although our conclusion is valid for  generic interacting  potentials, it  first emerged from a detailed study of the sine-Gordon model and in particular from the fact that solitonic  pseudo-spherical surfaces embedded in $\mathbb R^3$ are left invariant by the deformation. Analytic and numerical results concerning the perturbation of specific sine-Gordon soliton solutions are presented. 
}

\begin{document}
\maketitle
\flushbottom

\section{Introduction}
The deformation of 2D quantum field theories  \cite{Smirnov:2016lqw, Cavaglia:2016oda} by the Zamolodchikov's $\TbT$ operator  \cite{Zamolodchikov:2004ce},  has recently attracted the attention of theoretical physicists due to the many  important links with string theory \cite{Dubovsky:2012wk, dubovsky2012effective,Caselle:2013dra, Chen:2018keo}  and AdS/CFT \cite{McGough:2016lol,Turiaci:2017zwd, Giveon:2017nie, Giveon:2017myj,Asrat:2017tzd,Giribet:2017imm,Kraus:2018xrn, Cottrell:2018skz,  Baggio:2018gct,Babaro:2018cmq}. 

A remarkable  property of this perturbation, discovered  in \cite{Smirnov:2016lqw, Cavaglia:2016oda}, concerns the evolution of the quantum spectrum at finite volume $R$, with periodic boundary conditions,  in terms of the  $\TbT$ coupling constant $\tau$. The spectrum is governed by the inhomogeneous  Burgers  equation
\beq
\label{eq:Burgers}
\partial_{\ta} E_n(R,\ta) = \frac{1}{2} \partial_R\left( E_n^2(R,\ta) - P_n^2(R) \right) \;,
\eeq
where  $E_n(R,\ta)$ and $P_n(R)$ are the total energy and momentum of a generic energy eigenstate $|n \ket$, respectively. Equation  (\ref{eq:Burgers}) is valid  also for non-integrable  models. 

Notice that (\ref{eq:Burgers}) reveals an important feature of $\TbT$-deformed  QFTs: the interaction between the perturbing operator and the geometry, through the coupling  $\ta$. 
The latter property is a   basic requirement  for any sensible theory of gravity but  in the current case it  naturally emerges, non perturbatively and at full quantum level,  from a specific irrelevant perturbation of  Lorentz-invariant Quantum Field Theories (QFTs). An important link  with  JT topological gravity was noticed and studied in  \cite{Dubovsky:2017cnj}, where it  was shown
that  JT gravity coupled to matter leads to a scattering phase matching that associated to the $\TbT$ perturbation \cite{Mussardo:1999aj, Zamolodchikov:1991vx, Dubovsky:2012wk, dubovsky2012effective, Caselle:2013dra, Dubovsky:2013ira, Smirnov:2016lqw, Cavaglia:2016oda}.

Studies of   partition functions \cite{Cavaglia:2016oda, Cardy:2018sdv,  Dubovsky:2018bmo, Datta:2018thy} have led  to a proof of  the uniqueness of this perturbation  \cite{Aharony:2018bad} under the assumption that the theory on the torus is  invariant under modular transformations and that the energy of a given eigenstate is a function only of $\ta$ and of the energy and momentum of the corresponding state at $\ta= 0$. Furthermore,  starting from the  JT-gravity setup, in \cite{Dubovsky:2018bmo} the hydrodynamic-type equation (\ref{eq:Burgers}) was recovered. The latter   result together  with   \cite{Dubovsky:2017cnj} confirms,  beyond any reasonable doubt, the equivalence between the $\TbT$ deformation and JT topological gravity coupled to  generic matter field.

The aim of this paper is to address the problem concerning the classical interpretation of the $\TbT$ perturbation following the  more direct approach proposed  in \cite{Cavaglia:2016oda} and further developed in \cite{Bonelli:2018kik, Conti:2018jho}. The current analysis is  based on the observation \cite{Smirnov:2016lqw, Cavaglia:2016oda} that  (\ref{eq:Burgers}) directly implies a self-consistent  flow  equation for the deformed  Lagrangian  $\cL^{(\ta)}$
\beq
\label{eq:lag}
\partial_\ta \cL^{(\ta)} = \textrm{Det}\(T^{(\ta)}_{\mu \nu}\) \;,\quad   T^{(\ta)}_{\mu \nu} = -\fract{2}{\sqrt{|g|}} \fract{\delta  (\cL^{(\ta)} \sqrt{|g|})}{\delta g^{\mu \nu}}\;,
\eeq
where $g=\textrm{Det}\(g_{\mu\nu} \)$ and $\TbT= -\pi^2 \textrm{Det}\(T_{\mu \nu}\)$ is the   classical counterpart of Zamolodchikov's  operator.

Starting from the   unperturbed Lagrangian $\cL^{(0)}$   equation  (\ref{eq:lag}) can be solved  giving  the $\TbT$-deformed exact result  $\cL^{(\ta)}$. Adopting this strategy,  the Nambu-Goto classical Lagrangian  in the static gauge was recovered \cite{Cavaglia:2016oda} along with the deformation of bosonic models with generic interacting potential \cite{Cavaglia:2016oda,Bonelli:2018kik, Conti:2018jho}, WZW and $\sigma$-models \cite{Baggio:2018gct, Bonelli:2018kik, Dei:2018mfl}, and the  Thirring model \cite{Bonelli:2018kik}.

There are many reasons to study  these newly-discovered set of classical Lagrangians. First of all, according to  \cite{Dubovsky:2017cnj, Dubovsky:2018bmo}, these systems  should correspond to JT gravity coupled to non-topological matter, a fact  that is by no mean evident from the  Lagrangian point of view. 

Secondly, when the starting model is integrable, there should be a general way   to deform the whole integrable model machinery. For example, a generalisation of the  ODE/IM correspondence \cite{Dorey:1998pt, Bazhanov:1998wj,  Dorey:2007zx} should lead to an alternative method  to obtain  the quantum  spectrum at finite volume \cite{Lukyanov:2010rn, Dorey:2012bx} and it may  open the  way to the inclusion of the $\TbT$  inside the Wilson Loops/Scattering Amplitudes setup, in $\text{AdS}_5/\text{CFT}_4$ \cite{Alday:2009dv,Alday:2010vh} and perhaps also to consistently deform the Argyres-Douglas theory \cite{Gaiotto:2014bza,Ito:2017ypt, Grassi:2018bci}.  

The main purpose of this article is to  prove that, for bosonic   theories with arbitrary interacting potentials,  the $\TbT$ perturbation has indeed  the alternative interpretation as a space-time deformation. In Euclidean coordinates the change of variables is
\beqa
dx^\mu &=& \left(\delta^{\mu}_{\;\;\nu} + \ta \widetilde T^{\mu}_{\;\;\nu}(\mathbf y) \right) dy^\nu \;, \quad \mathbf y=(y^1,y^2)\;,  \\
dy^\mu &=& \left(\delta^{\mu}_{\;\;\nu} + \ta \bigl(\widetilde T^{(\tau)}\bigr)^{\mu}_{\;\;\nu}(\mathbf x) \right) dx^\nu \;, \quad \mathbf x=(x^1,x^2) \;,
\label{eq:map1}
\eeqa
with  $\widetilde T^{\mu}_{\;\;\nu} = -\epsilon^\mu_{\;\;\rho}\epsilon^\sigma_{\;\;\nu}T^\rho_{\;\;\sigma}$ and $\bigl(\widetilde T^{(\ta)}\bigr)^{\mu}_{\;\;\nu} = -\epsilon^\mu_{\;\;\rho}\epsilon^\sigma_{\;\;\nu}\bigl(T^{(\ta)}\bigr)^\rho_{\;\;\sigma}$,  where $T=T^{(0)}$ and $T^{(\ta)}$ are the unperturbed and perturbed stress-energy tensor in the set of coordinates $\mathbf y$ and $\mathbf x$, respectively. Then, any  solution of the perturbed EoMs can be mapped  onto the  $\ta=0$  corresponding solution, {\it i.e.}
\beq
\phi^{(\ta)}(\mathbf x ) = \phi^{(0)}\( \mathbf y( \mathbf x) \) \;,  
\label{eq:solutiontrans}
\eeq
where the r.h.s. of (\ref{eq:solutiontrans})\footnote{Notice that from (\ref{eq:solutiontrans}) it follows that $\phi^{(\ta)}(\mathbf x )$ fulfills the Burgers-type equation
\beq
\partial_\ta \phi^{(\ta)}({\bf x}) + (\partial_\ta x^{\mu}) \,\partial_\mu \phi^{(\ta)}({\bf x})=0 \;,
\eeq 
which may justify the wave-breaking phenomena observed in section \ref{sec:Num}. In our results $x^\mu$ is always linear in $\ta$, however we could not find an explicit expression for $\partial_\ta x^\mu$ valid in general.} is defined on a deformed space-time with metric
\beq
{\bf g}'_{\mu\nu} = \delta_{\mu\nu}-\ta\epsilon_{\mu\rho}\epsilon^\sigma_{\;\;\nu}\( 2T+\ta T^2 \)^\rho_{\;\;\sigma} \;.
\label{eq:metric1}
\eeq
In fact (\ref{eq:map1}) corresponds to a natural generalization of the Virasoro conditions used in the GGRT treatment of the NG string \cite{Goddard:1973qh}, \footnote{See \cite{McGough:2016lol} for a clarifying discussion related to the current topic.} and it matches precisely  the generalisation corresponding to classical JT gravity  \cite{Dubovsky:2017cnj, Dubovsky:2018bmo}.
\section{Classical integrable equations and embedded surfaces}
\label{sec:class}
It is an established fact that integrable equations in two dimensions admit an interpretation in terms of surfaces embedded inside an $N$-dimensional space. The two oldest examples of this connection, dating back to the works of 19th century geometers \cite{Bour_862, Liou_853}, are the sine-Gordon and Liouville equations. They appear as the Gauss-Mainardi-Codazzi (GMC) system of equations (\ref{eq:GMC}) for, respectively, pseudo-spherical and minimal surfaces embedded in the Euclidean space $\mathbb R^3$. As proved by Bonnet \cite{Bonn_867}, any surface embedded in $\mathbb R^3$ is uniquely determined (up to its position in the ambient space) by two rank $2$ symmetric tensors: the metric $g_{\mu\nu}$ (\ref{eq:metric_tensor}) and the second fundamental tensor $d_{\mu\nu}$ (\ref{eq:second_fund_tensor}). Their intuitive role is to measure, respectively, the length of an infinitesimal curve and the displacement of its endpoint from the tangent plane at the starting point. One can then use $g_{\mu\nu}$ and $d_{\mu\nu}$ to study the motion of a frame anchored to the surface. The result is a system of linear differential equations, known as \emph{Gauss-Weingarten} equations (\ref{eq:Gauss_eq_GWsystem}, \ref{eq:Weingarten_eq_GWsystem}). The GMC system appears then as the consistency condition for this linear system, effectively constraining the ``moduli space" consisting of the two tensors $g_{\mu\nu}$ and $d_{\mu\nu}$.

The search for a general correspondence originated in the works of Lund, Regge, Pohlmeyer and Getmanov \cite{Pohl_76,Lund_Regg_76,Getm_77} and was subsequently formalised by Sym \cite{Sym_82,Sym_83,Sym_84_1,Sym_84_2,Sym_84_3} who showed that any integrable system whose associated linear problem is based on a semi-simple Lie algebra $\mathfrak g$ can be put in the form of a GMC system for a surface embedded in a $\textrm{dim}(\mathfrak g)$-dimensional surface.\footnote{An interesting additional result of Sym concerns the existence of the same kind of connection for spin systems and $\sigma$-models.} In this section we will shortly review Sym's results for the general setting and concentrate on the case of sine-Gordon model. We will use the following conventions
\begin{equation*}
	\mathbf z = \left(z^1,z^2\right)\;,\quad \partial_\mu f\left(\mathbf z\right) \equiv \frac{\partial}{\partial z^{\mu}} f\left(\mathbf z\right) \;,\quad \forall f\;:\;\mathbb R^2\;\rightarrow \mathbb R\;,\quad \mu=1,2\;.
\end{equation*}

\subsection{Construction of the solitonic surfaces}
Let us consider a generic $2$-dimensional system of non-linear partial differential equations for a set of real fields $\left\lbrace\phi_i\left(\mathbf z\right)\right\rbrace$ admitting a \emph{Zero Curvature Representation} (ZCR) for a pair of functions $L_1$ and $L_2$ taking values in a $d$-dimensional representation of a semi-simple Lie algebra\footnote{Here we abuse notations by denoting with $\mathfrak g$ both the algebra and its $d$-dimensional representation. The same applies for the associated Lie Group $G$.} $\mathfrak g$ ($\textrm{dim}(\mathfrak g) = N$):
\begin{equation}
	\partial_2 L_1 - \partial_1 L_2 +\left[L_1,L_2\right] = 0\;.
\label{eq:ZCR}
\end{equation}
The functions $L_\mu$ depend on $\mathbf z$ through the fields $\phi_i\left(\mathbf z\right)$ and their derivatives and on a real spectral parameter $\lambda$:
\begin{equation}
	L_\mu \equiv L_\mu\left(\mathbf z\vert \lambda\right) \equiv L_\mu\left(\left\lbrace\phi_i\left(\mathbf z\right)\right\rbrace,\left\lbrace\partial_\nu\phi_i\left(\mathbf z\right)\right\rbrace, \ldots \vert \lambda\right)\;.
\end{equation}
The Zero Curvature Representation can be interpreted as the compatibility condition for a system of first-order linear partial differential equations involving an auxiliary $d\times d$ matrix-valued function $\Phi \equiv \Phi\left(\mathbf z\vert \lambda\right)$
\begin{equation}
\label{eq:LaxC}
	\partial_\mu \Phi = L_\mu \Phi\;,\quad \mu = 1,2\;,
\end{equation}
commonly known as \emph{associated linear problem}. Assuming $\Phi\left(\mathbf z_0\vert \lambda\right)\in G$ as initial condition, with $G$ being the Lie group associated to $\mathfrak g$, equation (\ref{eq:LaxC}) allows, in principle, to recover a single-valued function $\Phi\in G$ in the whole $\mathbb R^2$. This function can then be used to construct the following object
\begin{equation}
	r\left(\mathbf z\vert \lambda\right) = \Phi^{-1}\left(\mathbf z\vert \lambda\right)\frac{\partial}{\partial \lambda} \Phi\left(\mathbf z\vert \lambda\right)\;,
\label{eq:positionvectoring}
\end{equation}
which is interpreted as the coordinate description of a $\lambda$-family of surfaces embedded into the $N$-dimensional affine space $\mathfrak g$. Moreover, equipping the affine space $\mathfrak g$ with a non-degenerate scalar product (i.e. the Killing form of the semi-simple Lie algebra), we can convert $\mathfrak g$ into an $N$-dimensional flat space. In other words, we can find an orthonormal basis $\left\lbrace\mathbf e^i \right\rbrace$ of $\mathfrak g$ with respect to the Killing form and then extract the quantities $r_i$ from the identity
\begin{equation}
	r = \sum_{i=1}^N r_i \mathbf e^i = \Phi^{-1}\left(\mathbf z\vert \lambda\right)\frac{\partial}{\partial \lambda} \Phi\left(\mathbf z\vert \lambda\right)\;.
\label{eq:coordinate_vector}
\end{equation}
The vector $\mathbf r= \left(\begin{array}{c c c c} r_1, r_2, \ldots , r_N\end{array}\right)^T$ is then the position vector of a family of surfaces embedded in $N$-dimensional flat space,\footnote{The signature of this space depends on the real form chosen for the algebra; for example $\mathfrak{sl}\left(2\right) \simeq \mathfrak{so}\left(2,1\right)$ give rise to surfaces in Minkowski space $\mathbb R^{2,1}$.} parametrised by $\lambda$. These are called \emph{solitonic surfaces} and satisfy the following properties:
\begin{enumerate}
	\item their GMC system reduces to the ZCR (\ref{eq:ZCR}). This means that any integrable system whose EoMs can be represented as a ZCR depending on a spectral parameter $\lambda$, can be associated to a particular class of surfaces;
	\item they are invariant with respect to $\lambda$-independent gauge transformation of the pair $L_{\mu}$. This fact provides a way to prove the equivalence of distinct soliton systems up to gauge transformations and independent coordinate redefinitions, see \cite{Sym_84_2};
	\item their metric tensor (induced by the flat space $\mathfrak g$) is explicitly computed from the pair $L_{\mu}$ as
	\begin{equation}
		g_{\mu\nu} = \textrm{Tr} \left( \textrm{\bf Ad}\(\frac{\partial L_{\mu}}{\partial \lambda}\) \textrm{\bf Ad}\(\frac{\partial L_{\nu}}{\partial \lambda}\)\right)\;,
	\label{eq:metric_from_lax}
	\end{equation}
where $\textrm{\bf Ad}$ denotes the adjoint representation of the algebra $\mathfrak g$. Consequently, any intrinsic property of the soliton surface is determined uniquely by the ZCR.
\end{enumerate}
\subsection{The case of sine-Gordon}
Let us now consider the specific case of the sine-Gordon equation
\begin{equation}
	\partial \bar \partial \phi = \frac{m^2}{\beta} \sin\left(\beta \phi\right)\;,
\end{equation}
where we set $\mathbf z=(z^1,z^2)= (z, \bar z)$. The ZCR for this model is well known 
\begin{align}
	L_1^{\textrm{sG}} = \mathcal Z = \frac{\beta}{2}\partial\phi \; \cS^3 + \mathbbm{i} m\lambda \left[\cos\left(\frac{\beta}{2}\phi\right) \cS^1 - \sin\left(\frac{\beta}{2}\phi\right) \cS^2\right]\;, \label{eq:sG_ZCR_1}\\
	L_2^{\textrm{sG}} = \bar{\mathcal Z} = -\frac{\beta}{2}\bar \partial\phi \; \cS^3 + \mathbbm{i}\frac{m}{\lambda}\left[\cos\left(\frac{\beta}{2}\phi\right) \cS^1 + \sin\left(\frac{\beta}{2}\phi\right) \cS^2\right]\;, \label{eq:sG_ZCR_2}
\end{align}
where $\cS^j$ are the generators of $\mathfrak{su}\left(2\right)$
\begin{equation}
	\left[\cS^i,\cS^j\right] = \varepsilon^{ij}_{\phantom{ij}k}\;\cS^k\;.
\end{equation}
Since $\textrm{dim}\(\mathfrak{su}\left(2\right)\)=3$, we know that we are dealing with a surface embedded in the Euclidean plane $\mathbb R^3$ ($\mathfrak{su}\left(2\right)$ is compact). As mentioned in section \ref{sec:class}, Bonnet theorem \cite{Bonn_867} tells us that any surface in $\mathbb R^3$ is completely specified (modulo its position) by its first and second fundamental quadratic forms, which can be computed easily:\footnote{These can be recovered by plugging (\ref{eq:coordinate_vector}) in the classical geometry formulae $$g_{\mu\nu} = \partial_\mu \mathbf r\cdot\partial_\nu \mathbf r\;,\quad d_{\mu\nu} = -\partial_\mu \partial_\nu  \mathbf r \cdot \mathbf n \;,$$ where $\mathbf n$ is the normal unit vector to the plane spanned by $\partial_1 \mathbf r$ and $\partial_2 \mathbf r$: $$\mathbf n = \frac{\partial_1 \mathbf r\times \partial_2 \mathbf r}{\left\vert \partial_1 \mathbf r\times \partial_2 \mathbf r\right\vert}\;.$$}
\begin{align}
	\textsc I^{\textrm{sG}} & = g^{\textrm{sG}}_{\mu \nu}dz^{\mu} dz^{\nu} = 2m^2\left[ \left(dz\right)^2 -\frac{2}{\lambda^2}\cos\left(\beta\phi\right) dz d\bar z + \frac{1}{\lambda^4} \left(d\bar z\right)^2 \right]\;, \label{eq:I_sG}\\
	\textsc{II}^{\textrm{sG}} & = d^{\textrm{sG}}_{\mu\nu}dz^{\mu}dz^{\nu} = 2m^2\frac{\sqrt{2}}{\lambda} \sin\left(\beta\phi\right) dz d\bar z \;. \label{eq:II_sG}
\end{align}
From (\ref{eq:I_sG}) and {\ref{eq:II_sG}) one can then extract the Gaussian and the mean curvatures using  (\ref{eq:Gauss_mean_curv_def}):
\begin{equation}
\label{eq:KHsG}
	 K^{\textrm{sG}} =\textrm{Det}\(d^{\textrm{sG}}_{\mu\rho}\,\left(g^{\textrm{sG}}\right)^{\rho\nu}\) = -\frac{\lambda^2}{2}\;,\quad  H^{\textrm{sG}} = d^{\textrm{sG}}_{\mu\nu}\,\left(g^{\textrm{sG}}\right)^{\nu\mu} =  \frac{\lambda}{\sqrt{2}}\cot\left(\beta\phi\right)\;,
\end{equation}
with $g^{\textrm{sG}}_{\mu\nu}\left(g^{\textrm{sG}}\right)^{\nu\rho} = \delta_{\mu}^{\rho}$. The fact that $ K^{\textrm{sG}}$ is constant negative tells us that we are dealing with a pseudo-spherical surface, which we were expecting from the old results of Bour \cite{Bour_862}. Thus, for this specific case, the solitonic surfaces correspond to pseudo-spherical ones, with the spectral parameter $\lambda$ playing the role of Gaussian curvature.
\section{The $\TbT$-deformed sine-Gordon model and its associated surfaces}
Let us now apply the Sym formalism sketched above to the $\TbT$-deformed sine-Gordon model \cite{Conti:2018jho}
\begin{align}
	&\partial\left(\frac{\bar\partial\phi}{S}\right) + \bar\partial\left(\frac{\partial\phi}{S}\right) = \frac{V'}{4S}\left(\frac{S+1}{1-\tau V}\right)^2\;, \label{eq:TTbarEom}\\
	&S = \sqrt{1+4\tau\left(1-\tau V\right)\partial\phi\bar\partial\phi}\;, \label{eq:TTbarS} \\
	&V = 2\frac{m^2}{\beta^2}\left(1-\cos(\beta\phi)\right)\;,\quad V' = 2\frac{m^2}{\beta}\sin(\beta \phi)\;,
\end{align}
and derive the geometric properties of the associated surfaces. We start with the ZCR, which was found in \cite{Conti:2018jho}
\begin{align}
	L_1^{\TbT} &\equiv \mathcal Z = \beta\frac{\partial\phi}{2S}\; \cS^3 + 2\mathbbm{i} m \left[F_+ \cos\left(\frac{\beta}{2}\phi\right) \cS^1 - F_- \sin\left(\frac{\beta}{2}\phi\right) \cS^2\right]\;, \label{eq:ZCR_TTbar_1}\\
	L_2^{\TbT} &\equiv \bar{\mathcal Z} = -\beta\frac{\bar \partial\phi}{2S} \; \cS^3 + 2\mathbbm{i} m \left[\bar F_+ \cos\left(\frac{\beta}{2}\phi\right) \cS^1+ \bar F_- \sin\left(\frac{\beta}{2}\phi\right) \cS^2\right]\;, \label{eq:ZCR_TTbar_2}
\end{align}
where
\begin{align}
	F_+ = \left(\lambda B_+ + \frac{1}{\lambda}\left(\partial\phi\right)^2 B_- \right)\;,\qquad F_- = \left(\lambda B_+ - \frac{1}{\lambda}\left(\partial\phi\right)^2 B_- \right)\;, \\
	F_- = \left(\frac{1}{\lambda} B_+ + \lambda\left(\bar \partial\phi\right)^2 B_- \right)\;,\qquad \bar F_- = \left(\frac{1}{\lambda} B_+ - \lambda\left(\bar \partial\phi\right)^2 B_- \right)\;,
\end{align}
with
\begin{equation}
    B_+ = \frac{\left(S+1\right)^2}{8S\left(1-\tau V\right)}\;,\qquad B_- = \frac{\tau}{2S}\;.
\end{equation}

Again we have a ZCR based on the algebra $\mathfrak{su}\left(2\right)$ and thus a surface embedded in $\mathbb R^3$. We need then to recover the fundamental forms I and II, whose computation, although straightforward as in the case of sine-Gordon, is lengthy and cumbersome. Sparing the uninteresting details, we present directly the results
\begin{align}
	\textrm{I}^{\TbT} &= g^{\TbT}_{\mu\nu}dz^{\mu}dz^{\nu} = \frac{m^2}{2S^2}\left(\frac{S+1}{1-\tau V}\right)^2\hat g_{\mu\nu}dz^{\mu}dz^{\nu}\;, \label{eq:I_TTbar}\\
	\textrm{II}^{\TbT} &= d^{\TbT}_{\mu\nu}dz^{\mu}dz^{\nu} = \frac{m^2\sin\left(\beta\phi\right)}{\sqrt{2}\lambda\left(1-\tau V\right)} \left(\frac{S+1}{S}\right)^2 \hat d_{\mu\nu}dz^{\mu}dz^{\nu}\;, \label{eq:II_TTbar}
\end{align}
where the matrices $\hat g_{\mu\nu}$ and $\hat d_{\mu\nu}$ are
\begin{equation}
	\hat g_{\mu\nu} = \left(\begin{array}{c c}
					\left(\frac{S+1}{2} - \frac{S-1}{2\lambda^2}\frac{\partial\phi}{\bar\partial\phi}\right)^2 + \frac{S^2-1}{4\lambda^2}\frac{\beta^2 V}{m^2} \frac{\partial\phi}{\bar\partial\phi} &  \frac{S^2-1}{4}\left(\frac{\bar\partial\phi}{\partial\phi}+\frac{1}{\lambda^4}\frac{\partial\phi}{\bar\partial\phi}\right) -\frac{S^2+1}{2\lambda^2}\cos\left(\beta\phi\right) \\
					\frac{S^2-1}{4}\left(\frac{\bar\partial\phi}{\partial\phi}+\frac{1}{\lambda^4}\frac{\partial\phi}{\bar\partial\phi}\right) -\frac{S^2+1}{2\lambda^2}\cos\left(\beta\phi\right) & \left(\frac{S+1}{2\lambda^2} - \frac{S-1}{2}\frac{\bar\partial\phi}{\partial\phi}\right)^2 + \frac{S^2-1}{4\lambda^2}\frac{\beta^2 V}{m^2} \frac{\bar\partial\phi}{\partial\phi}
				\end{array}\right)_{\mu\nu} \;, \notag
\label{eq:g_TTbar}
\end{equation}
\begin{equation}
	\hat d_{\mu\nu} = \left(\begin{array}{c c}
					\tau\left(\partial\phi\right)^2 & \frac{S^2+1}{4\left(1-\tau V\right)} \\
					\frac{S^2+1}{4\left(1-\tau V\right)} & \tau\left(\bar\partial\phi\right)^2
				\end{array}\right)_{\mu\nu} \;.
\label{eq:d_TTbar}
\end{equation}
One easily verifies that in the $\tau\rightarrow 0$ limit, which implies $S\rightarrow 1$, one recovers the fundamental forms of sine-Gordon
\begin{align}
	\textrm{I}^{\TbT} & \underset{\tau\rightarrow 0}{\rightarrow} 2 m^2 \left(\begin{array}{c c}
		1 & -\frac{1}{\lambda^2}\cos\left(\beta\phi\right)   \\
		-\frac{1}{\lambda^2}\cos\left(\beta\phi\right) & \frac{1}{\lambda^4}
	\end{array}\right)_{\mu\nu}dz^{\mu}dz^{\nu} = \textrm{I}^{\textrm{sG}}\;, \\
	\textrm{II}^{\TbT} & \underset{\tau\rightarrow 0}{\rightarrow} m^2\frac{\sqrt{2}}{\lambda} \sin\left(\beta\phi\right) \left(\begin{array}{c c}
		0 & 1 \\
		1 & 0
	\end{array}\right)_{\mu\nu}dz^{\mu}dz^{\nu} = \textrm{II}^{\textrm{sG}}\;.
\end{align}
What is striking about the matrices (\ref{eq:g_TTbar}) is that, although their dependence on $\tau$ is complicated, they recombine in such a way that the Gaussian and mean curvature do not depend explicitly on it! In fact these two geometric invariants  are exactly the same as the unperturbed sine-Gordon model:
\begin{equation}
\label{eq:KHTT}
	 K^{\TbT} = -\frac{\lambda^2}{2}=  K^{\textrm{sG}} \;, \quad  H^{\TbT} = \frac{\lambda}{\sqrt{2}}  \cot\left(\beta\phi\right)=  H^{\textrm{sG}} \;.
\end{equation}
This suggests that the solitonic surface corresponding to a particular
solution of the $\TbT$-deformed sine-Gordon equation is the same as the one associated to the undeformed model, what changes should be the coordinate system used to describe it. % 
For the sake of completeness, we have reported  in Figure \ref{fig:surfaces} examples of embedded pseudo-spherical surfaces related to  one-kink solutions, a stationary breather and a two-kink solution. The plots were obtained implementing the method described in \cite{rogers2002backlund}.   
\begin{figure}[h]
  \centering
  \subfloat[]{\includegraphics[scale=0.19]{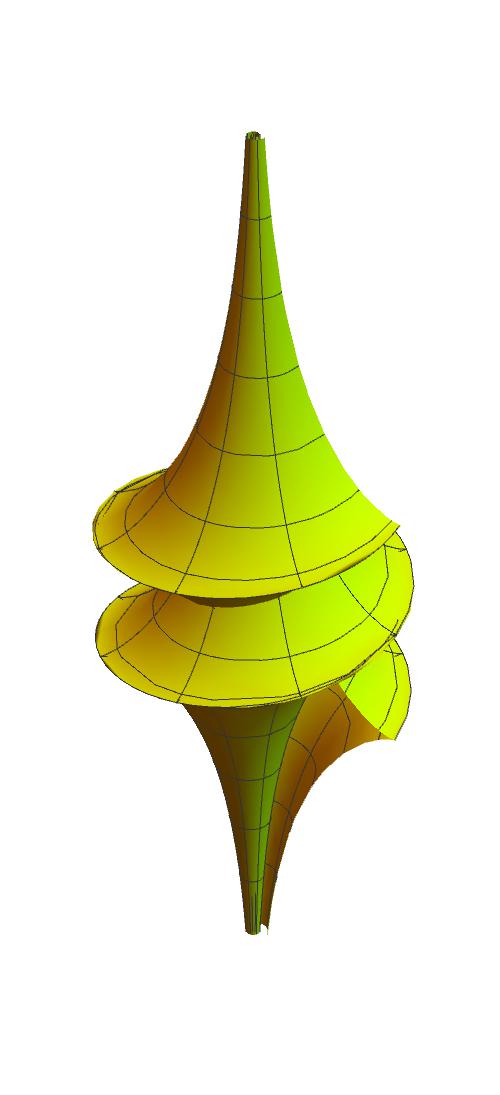}
  \label{fig:Dini}}
%  \hspace{3.5cm}
  \subfloat[]{\includegraphics[scale=0.19]{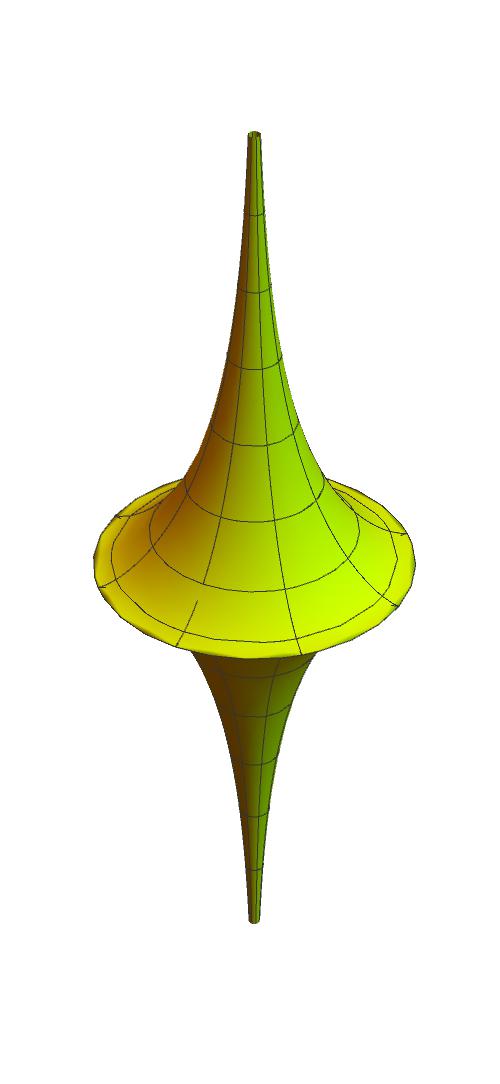}
  \label{fig:Beltrami}}
%  \vspace{0.2cm}
  \subfloat[]{\includegraphics[scale=0.28]{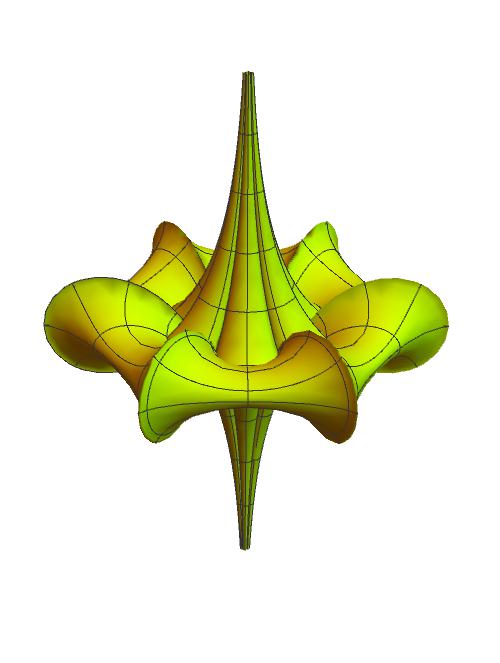}
  \label{fig:BreatherS}}
   \hspace{0cm}
  \subfloat[]{\includegraphics[scale=0.23]{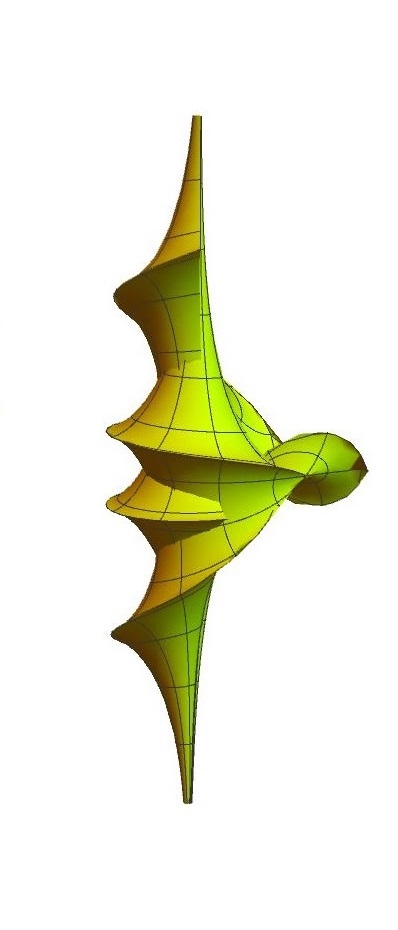}
  \label{fig:2kinkS}}
%  \vspace{0.1cm}
%
\caption{ Pseudo-spherical solitonic surfaces associated to  kink  and breather solutions. Figure  \ref{fig:Dini}  represents the Dini surface, corresponding to a moving kink, while in Figure \ref{fig:Beltrami}  the famous Beltrami pseudo-sphere is represented. The latter surface is obtained from  Dini's surface  by taking the stationary  limit of the kink solution. Figures \ref{fig:BreatherS} and \ref{fig:2kinkS} correspond  to the  pseudo-spherical surfaces associated to a stationary  breather  and to a two-kink solution, respectively.}
\label{fig:surfaces}
\end{figure}
The  embedded surfaces in $\RR^3$, as we have just argued and  will be explicitly shown in the next section, are independent of the deformation parameter $\ta$, being it 
re absorbable through a local change of coordinates.

The corresponding soliton solutions, described  in section \ref{sec:Num}, are instead affected by  the $\TbT$  in a highly non-trivial way. For instance, they  generally possess critical values in $\ta$ corresponding to shock-wave phenomena, {\it i.e.} branching of the solutions. Examples of shock-wave phenomena and square root-type transitions in the classical energy -- similar to the Hagedorn transition at quantum level -- will be discussed in sections \ref{sec:Num} and \ref{sec:Hagedorn} for specific solutions of the deformed sine-Gordon model.

\subsection{From the  deformed to the  undeformed model through a local change of coordinates}
\label{sec:change}
Thus we have inferred that there must exist a coordinate system $\mathbf w=( w^1(\mathbf z), w^2(\mathbf z))=\left(w(\mathbf z),\bar w(\mathbf z)\right)$ in which the matrices $g_{\mu\nu}^{\TbT}$ and $d_{\mu\nu}^{\TbT}$ assume the same form as  $g_{\mu\nu}^{\textrm{sG}}$ and $d_{\mu\nu}^{\textrm{sG}}$, respectively. In formulae
\begin{align}
	g_{\mu\nu}^{\textrm{sG}}dw^{\mu}dw^{\nu} = g_{\mu\nu}^{\TbT}dz^{\mu}dz^{\nu}\;\Longrightarrow\;g_{\mu\nu}^{\textrm{sG}}\frac{dw^{\mu}}{dz^{\rho}}\frac{dw^{\nu}}{dz^{\sigma}} = g_{\rho\sigma}^{\TbT}\;, \\
	d_{\mu\nu}^{\textrm{sG}}dw^{\mu}dw^{\nu} = d_{\mu\nu}^{\TbT}dz^{\mu}dz^{\nu}\;\Longrightarrow\;d_{\mu\nu}^{\textrm{sG}}\frac{dw^{\mu}}{dz^{\rho}}\frac{dw^{\nu}}{dz^{\sigma}} = d_{\rho\sigma}^{\TbT}\;.
\end{align}
It is now a matter of simple algebraic manipulations to obtain the following equations for the new coordinates
\begin{align}
	\partial w &= \frac{\left(S+1\right)^2}{4S\left(1-\tau V\right)}\;,\qquad \bar\partial \bar w = \frac{\left(S+1\right)^2}{4S\left(1-\tau V\right)}\;, \label{eq:geometric_map_1}\\
	\bar\partial w &= \frac{\tau}{S} \left(\bar\partial\phi\right)^2\;,\qquad \quad\,\;\;\partial\bar w = \frac{\tau}{S}\left(\partial\phi\right)^2\;. \label{eq:geometric_map_2}
\end{align}
Let us now use the latter relations to find the partial derivatives of the field $\phi$ in the coordinates $\mathbf w$:
\begin{equation}
	\left(\begin{array}{c} \partial\phi \\ \bar\partial\phi \end{array}\right) = \mathcal J \left( \begin{array}{c}  \partial\phi / \partial w \\ \partial\phi / \partial\bar w \end{array}\right)\;,\qquad \mathcal J = \left(\begin{array}{c c}
		\partial w & \partial\bar w \\
		\bar \partial w & \bar\partial\bar w
	\end{array}\right)\;.
\end{equation}
The result is
\begin{align}
		\partial \phi = \frac{1}{1-\tau \left(\mathcal K + V\right)}\frac{\partial \phi}{\partial w}\;,\qquad \bar \partial\phi = \frac{1}{1-\tau \left(\mathcal K + V\right)}\frac{\partial \phi}{\partial\bar w}\;,
\label{eq:geometric_map_phi_sG}
\end{align}
where we have defined the following function
\begin{equation}
	\mathcal K = \frac{\partial\phi(\mathbf w)}{\partial w}\frac{\partial\phi(\mathbf w)}{\partial\bar w} \;.
\end{equation}
With the help of (\ref{eq:geometric_map_phi_sG}), we can now find the expression for $S$ in the coordinates  $\mathbf w$
\begin{equation}
	S = \sqrt{1+4\tau\left(1-\tau V\right) \partial\phi \bar\partial\phi} = \frac{1+\tau  \left(\mathcal K - V\right)}{1-\tau \left(\mathcal K + V\right)}\;.
\label{eq:S_in_zeta_coord}
\end{equation}
We can then write the Jacobian matrix $\mathcal J$ and its inverse $\mathcal J^{-1}$ in terms of $\mathbf w$ as
\beqa
\label{eq:Jacobian}
	\mathcal J &=& \left(\begin{array}{c c}
		\partial w & \partial\bar w \\
		\bar \partial w & \bar\partial\bar w
	\end{array}\right) = \frac{1}{\left(1-\tau V\right)^2 - \tau^2 \mathcal K^2}  \left(\begin{array}{c c}
		1-\tau V & \tau \left(\frac{\partial \phi}{\partial  w}\right)^2 \\
		\tau \left(\frac{\partial \phi}{\partial \bar w}\right)^2 & 1- \tau V
	\end{array}\right)\;, \notag \\
		\mathcal J^{-1} &=& \left(\begin{array}{c c}
		\partial_ w z & \partial_{ w}\bar z \\
		\partial_{\bar w} z & \partial_{\bar w} \bar z
	\end{array}\right) =  \left(\begin{array}{c c}
		1-\tau V & -\tau \left(\frac{\partial \phi}{\partial  w}\right)^2 \\
		-\tau \left(\frac{\partial \phi}{\partial \bar w}\right)^2 & 1- \tau V
	\end{array}\right)\;.
\eeqa
This results allows us to express the partial derivatives of any function $f\(\mathbf z\)$ as partial derivatives with respect to the new coordinates
\begin{equation}
	\left(\begin{array}{c} \partial f \\ \bar\partial f \end{array}\right) = \mathcal J \left( \begin{array}{c} \partial f / \partial w \\ \partial f / \partial\bar w \end{array}\right)\;,
\end{equation}
and we can then apply all the above formulae to the equation (\ref{eq:TTbarEom}), obtaining
\begin{align}
	\partial\left(\frac{\bar \partial \phi}{S}\right) + \bar \partial\left(\frac{ \partial \phi}{S}\right) &= 2\frac{\frac{\partial}{\partial w}\frac{\partial}{\partial \bar w} \phi}{\left(1+\tau \left(\mathcal K -V\right)\right)^2} - 2 \tau \frac{V'}{\left(1-\tau V\right)^2-\tau^2\mathcal K^2} \frac{\mathcal K}{\left(1+\tau \left(\mathcal K -V\right)\right)}\;,
	\label{eq:equ1}\\
	\frac{V'}{4S}\left(\frac{S+1}{1-\tau V}\right)^2 &= \frac{V'}{\left(1-\tau V\right)^2-\tau^2\mathcal K^2}\;.
	\label{eq:equ2}
\end{align}
The equality of (\ref{eq:equ1}) and (\ref{eq:equ2}) yields then
\begin{equation}
	\frac{2\frac{\partial}{\partial w}\frac{\partial}{\partial \bar w} \phi- V'}{\left(1+\tau \left(\mathcal K -V\right)\right)^2} = 0\;. \label{eq:TTbarEoM_zeta_variables}
\end{equation}
\section{A geometric map for N-boson fields and arbitrary  potential}
We have seen, in the preceding section, how the $\TbT$ deformation of the sine-Gordon model can be interpreted as a field-dependent coordinate transformation. We arrived at this interesting conclusion by exploiting the relation existing amongst ZCR of soliton equations and the classical geometry of surfaces embedded in flat space. Although this connection was pivotal in guiding us to the map (\ref{eq:geometric_map_1}, \ref{eq:geometric_map_2}), from that point on we did not make explicit mention to the form of the potential. In other words, we can consider all formulae from (\ref{eq:geometric_map_1}) to  (\ref{eq:TTbarEoM_zeta_variables}) to be valid for any $2$-dimensional single scalar system.

More generally,  the results (\ref{eq:Jacobian}, \ref{eq:TTbarEoM_zeta_variables}) admit a straightforward  generalisation to the case of $N$ bosonic fields $\phi_i$, $(i=1,\dots,N)$ interacting with a generic derivative-independent potential $V(\phi_i)$
\beqa
\label{eq:TTbar_generic_N_scalars}
\mathcal{L}^{(\ta)}_N\( \mathbf z \) = \frac{V}{1-\ta V}+\frac{-1+\sqrt{1+4\bar\ta\(\mathcal{L}^{(0)}_{\text{free}}-\bar\ta\mathcal{B}\)}}{2\bar\ta} \;,\\
\quad \mathcal{L}^{(0)}_{\text{free}} = \sum_{i=1}^N\partial\phi_i\bar\partial\phi_i \;,\quad \mathcal{B} = | \partial \vec{\phi} \times \bar{\partial} \vec{\phi} |^2 \;, 
\eeqa
with $\bar\ta = \ta(1-\ta V)$, arising as a $\TbT$-deformation of \cite{Conti:2018jho,Bonelli:2018kik}
\beq
\mathcal{L}^{(0)}_N = \sum_{i=1}^N \partial\phi_i\bar\partial\phi_i + V(\phi_i) \;.
\eeq
The generalization of (\ref{eq:Jacobian}) to the $N$-boson case is
\beqa
\label{eq:JacobianN}
\mathcal{J}_N &=& \(\begin{array}{cc}
\partial w & \partial\bar w \\
\bar\partial  w & \bar\partial\bar w
\end{array}\) = 
\frac{1}{(1-\ta V)^2-\ta^2\(\mathcal{K}_N\)^2}\(\begin{array}{cc}
1-\ta V & \ta \sum_{i=1}^N\( \frac{\partial\phi_i}{\partial w} \)^2 \\
\ta \sum_{i=1}^N \( \frac{\partial\phi_i}{\partial\bar{ w}} \)^2 & 1-\ta V 
\end{array}\) \;, \notag \\
\mathcal{J}_N^{-1} &=& \(\begin{array}{cc}
\partial_ w z & \partial_ w \bar{z} \\
\partial_{\bar w} z & \partial_{\bar w} \bar z
\end{array}\) =
\(\begin{array}{cc}
1-\ta V & -\ta \sum_{i=1}^N\( \frac{\partial\phi_i}{\partial w} \)^2 \\
-\ta \sum_{i=1}^N \( \frac{\partial\phi_i}{\partial\bar{ w}} \)^2 & 1-\ta V 
\end{array}\) \;,
\eeqa
with $\(\mathcal{K}_N\)^2 = \sum_{i=1}^N\( \frac{\partial\phi_i}{\partial w} \)^2\sum_{i=1}^N\( \frac{\partial\phi_i}{\partial\bar w} \)^2$. In fact we have verified that the deformed EoMs resulting from (\ref{eq:TTbar_generic_N_scalars}) are mapped by (\ref{eq:JacobianN}) into the undeformed EoMs associated to $\mathcal{L}^{(0)}_N$.\\
It is instructive to translate (\ref{eq:JacobianN}) in Euclidean coordinates. Considering
\beqa
\(\frac{\partial}{\partial w}+\frac{\partial}{\partial\bar w}\)(z+\bar{z}) &=& 2 + \ta\( -\sum_{i=1}^N\( \frac{\partial\phi_i}{\partial\bar{ w}} \)^2 - \sum_{i=1}^N\( \frac{\partial\phi_i}{\partial w} \)^2 - 2V \) \;, \notag \\
\(\frac{\partial}{\partial w}-\frac{\partial}{\partial\bar w}\)(z-\bar{z}) &=& 2 - \ta \(\sum_{i=1}^N\( \frac{\partial\phi_i}{\partial\bar{ w}} \)^2 - \sum_{i=1}^N\( \frac{\partial\phi_i}{\partial w} \)^2 + 2V \) \;, \notag \\
\(\frac{\partial}{\partial w}+\frac{\partial}{\partial\bar w}\)(z-\bar{z}) &=& \ta\( \sum_{i=1}^N\( \frac{\partial\phi_i}{\partial\bar{ w}} \)^2 - \sum_{i=1}^N\( \frac{\partial\phi_i}{\partial w} \)^2 \) \;, \notag \\
\(\frac{\partial}{\partial w}-\frac{\partial}{\partial\bar w}\)(z+\bar{z}) &=& -\ta\( \sum_{i=1}^N\( \frac{\partial\phi_i}{\partial\bar{ w}} \)^2 - \sum_{i=1}^N\( \frac{\partial\phi_i}{\partial w} \)^2 \) \;, 
\eeqa
and moving to Euclidean coordinates both in the $\mathbf z$ and in the $\mathbf w$ frames
\beq
\begin{cases}
z = x^1+\mathbbm{i}\,x^2 \\
\bar{z} = x^1-\mathbbm{i}\,x^2
\end{cases} \;,\;
\begin{cases}
 w = y^1+\mathbbm{i}\,y^2 \\
\bar{ w} = y^1-\mathbbm{i}\,y^2
\end{cases}\rightarrow
\begin{cases}
\frac{\partial}{\partial w}+\frac{\partial}{\partial\bar w} = \frac{\partial}{\partial y^1} \\
\frac{\partial}{\partial w}-\frac{\partial}{\partial\bar w} = -\mathbbm{i}\frac{\partial}{\partial y^2}
\end{cases}
\eeq
we find
\beq
\label{eq:invjac_comp}
\frac{\partial x^1}{\partial y^1} = 1+\ta T^2_{\;\;2}(\mathbf y) \;,\quad
\frac{\partial x^2}{\partial y^2} = 1+\ta T^1_{\;\;1}(\mathbf y) \;,\quad
\frac{\partial x^1}{\partial y^2} = \frac{\partial x^2}{\partial y^1} = -\ta T^1_{\;\;2}(\mathbf y) \;,
\eeq
where $T^\mu_{\;\;\nu}(\mathbf y)$ is the stress energy tensor of the undeformed theory, $T=T^{(0)}$. Expressions (\ref{eq:invjac_comp}) can be more compactly rewritten as
\beq
\label{eq:invjac}
\frac{\partial x^\mu}{\partial y^\nu} = \delta^{\mu}_{\;\;\nu} + \ta \widetilde T^{\mu}_{\;\;\nu}(\mathbf y) \;,\quad \widetilde T^{\mu}_{\;\;\nu}(\mathbf y) = -\epsilon^\mu_{\;\;\rho}\epsilon^\sigma_{\;\;\nu}T^\rho_{\;\;\sigma}(\mathbf y) \;.
\eeq
From (\ref{eq:invjac}) the inverse Jacobian in Euclidean coordinates reads
\beq
\mathcal{J}_N^{-1} = 
\(\begin{array}{cc}
\frac{\partial x^1}{\partial y^1} & \frac{\partial x^2}{\partial y^1} \\
\frac{\partial x^1}{\partial y^2} & \frac{\partial x^2}{\partial y^2}
\end{array}\) =
\(\begin{array}{cc}
1+\ta T^2_{\;\;2} & -\ta T^1_{\;\;2} \\
-\ta T^1_{\;\;2} & 1+\ta T^2_{\;\;2}
\end{array}\) \;,
\eeq
and thus the metric, in the set of coordinates $\mathbf{y}$, is
\beq
{\bf g}'_{\mu\nu} = \frac{\partial x^\rho}{\partial y^\mu} \frac{\partial x^\sigma}{\partial y^\nu} {\bf g}_{\rho\sigma} = \delta_{\mu\nu}-\ta\epsilon_{\mu\rho}\epsilon^\sigma_{\;\;\nu}\( 2T+\ta T^2 \)^\rho_{\;\;\sigma} \;,
\eeq
where we used the fact that ${\bf g}_{\rho\sigma} = \delta_{\rho\sigma}$. Translating the first expression of (\ref{eq:JacobianN}) in $\mathbf z$ coordinates and then moving to Euclidean coordinates, one obtains the inverse relation of (\ref{eq:invjac})
\beq
\frac{\partial y^\mu}{\partial x^\nu} = \delta^{\mu}_{\;\;\nu} + \ta \bigl(\widetilde T^{(\tau)}\bigr)^{\mu}_{\;\;\nu}(\mathbf x) \;, \quad \bigl(\widetilde T^{(\tau)}\bigr)^{\mu}_{\;\;\nu}(\mathbf x) = -\epsilon^\mu_{\;\;\rho}\epsilon^\sigma_{\;\;\nu}\bigl(T^{(\tau)}\bigr)^{\mu}_{\;\;\nu}(\mathbf x) \;,
\eeq
where $\bigl(T^{(\tau)}\bigr)^{\mu}_{\;\;\nu}(\mathbf x)$ is the stress energy tensor of the deformed theory. \\
Finally let us conclude this section with a couple of remarks:
\begin{itemize}
    \item Consider the transformation of the Lagrangian\footnote{In the $N=1$ case, the transformed Lagrangian takes an even simpler expression
   \beq
    \mathcal L_{1}^{(\tau)}\(\mathbf z( \mathbf w) \) = \frac{\mathcal L_1^{(0)}\left( \mathbf w \right)}{1 - \tau \mathcal L_1^{(0)} ( \mathbf w )}\;.
    \eeq} (\ref{eq:TTbar_generic_N_scalars}) under the on-shell map (\ref{eq:JacobianN})
    \beq
    \mathcal L_N^{(\tau)}\(\mathbf z( \mathbf w) \) = \frac{\mathcal L_N^{(0)}(\mathbf w) + \ta\( \(\mathcal{K}_N\)^2 - V^2\)}{1 - 2\ta V -\ta^2\( \(\mathcal{K}_N\)^2-V^2 \)} \;.
    \label{eq:transformed_lagrangian}
    \eeq
    Using the latter expression together with  
    \beq
    \textrm{Det}\( \mathcal{J}_N^{-1} \) = \textrm{Det}\( \mathcal{J}_N \)^{-1} = 1 - 2\ta V -\ta^2\( \(\mathcal{K}_N\)^2-V^2 \) \;, 
    \eeq
    we find that the action transforms as
    \beqa
    \label{eq:action}
    \mathcal{A}\,[\phi] &=& \int dz\,d\bar{z}\,\mathcal{L}_N^{(\tau)}(\mathbf z ) = \int d w\,d\bar w\,\left|\textrm{Det}\(\mathcal{J}_N^{-1}\)\right|\,\mathcal{L}_N^{(\tau)}\(\mathbf z(\mathbf w)\)  \notag \\
    &=&  \int d w\,d\bar w\,\( \mathcal L_N^{(0)}( \mathbf w)+\ta\,\TbT^{(0)}( \mathbf w)\right) \;,
    \eeqa
    where $\TbT^{(0)}( \mathbf w) = \(\mathcal{K}_N\)^2-V^2$. Thus, we conclude that the action is not invariant under the change of variables. This is not totally surprising since the map (\ref{eq:JacobianN}) is on-shell, however it is remarkable that the (bare) perturbing field can be so easily identified once the change of variables is performed. Again, our result matches with \cite{Dubovsky:2017cnj}, where the $\TbT^{(0)}$ term emerges as a JT gravity contribution to the action.
    \item Notice that the EoMs associated to (\ref{eq:TTbar_generic_N_scalars}) for a generic potential $V$ are invariant under the transformation\footnote{We thank Sergei Dubovsky for questioning us about the possible existence of such symmetry of the EoMs.}
    \beq
    \label{eq:transfEoMs}
    \mathbf z\rightarrow \gamma\,\mathbf z \;,\quad \ta\rightarrow \gamma\,\ta \;,\quad V\rightarrow V-c \;,
    \eeq
    with $c$ constant and $\gamma = 1/(1-\ta c)$, which corresponds to the following change of variables at the level of the solutions
    \beq
    \label{eq:transfEoMs2}
    \phi_i^{(\ta)}\(\mathbf z \)\big|_V = \phi_i^{\(\gamma\,\ta\)}\(\gamma \mathbf z \)\big|_{V-c} \;,\quad i=1,\dots N \;,
    \eeq
    where the notation $\phi_i^{(\ta)}\big|_V$ means that $\phi_i^{(\ta)}$ is solution to the deformed EoMs with potential $V$.
\end{itemize}

\section{$\TbT$-deformed soliton solutions in the sine-Gordon model}
\label{sec:Num}
In this section we show how to compute $\textrm{T}\bar{\textrm{T}}$-deformed solutions of the sG model by explicitly evaluating the change of variables on specific solutions $\phi(\mathbf  w)$ of the undeformed theory. The idea is to solve the following sets of differential equations derived from the inverse Jacobian (\ref{eq:Jacobian})
\beq
\label{eq:diffmap}
\begin{cases}
\frac{\partial z(\mathbf w)}{\partial w} = 1-\ta V(\phi(\mathbf w)) \\
\frac{\partial z(\mathbf w)}{\partial\bar{ w}} = -\ta\( \frac{\partial\phi(\mathbf w)}{\partial\bar{ w}} \)^2
\end{cases} \;,\;
\begin{cases}
\frac{\partial\bar{z}(\mathbf w)}{\partial w} = -\ta\( \frac{\partial\phi(\mathbf w)}{\partial w} \)^2 \\
\frac{\partial\bar{z}(\mathbf w)}{\partial\bar{ w}} = 1-\ta V(\phi(\mathbf w))
\end{cases} \;,
\eeq
for $z(\mathbf w)$ and $\bar{z}(\mathbf w)$. Then from the inverse map, {\it i.e.} $ \mathbf w(\mathbf z)$, we evaluate the expression of the deformed solution as
\beq
\phi^{(\ta)}(\mathbf z )=\phi^{(0)}( \mathbf w(\mathbf z)) \;.
\eeq
In the following we will deal only with some of the simplest solutions of the sG model. In principle our approach applies for all the solutions, although we could not find an explicit result for the integrated map in the cases involving more than two solitons.\\
For sake of clarity, the computations shown in the following sections will be carried on in light cone coordinates, {\it i.e.} $(z,\bar z)$ and $(w,\bar w)$, however the plots will be displayed using space and time coordinates $(x,t)=\(x^1,x^2\)$.

\subsection{The one-kink solution}

Let us start with the one-kink solution moving with velocity $v$
\beq
\label{eq:1kink}
\phi_{1\text{-kink}}^{(0)}(\mathbf w) = 4 \arctan\( e^{\frac{m}{\beta}\( a w+\frac{1}{a}\bar{ w} \)} \) \;,\; a = \sqrt{\frac{1-v}{1+v}} \;.
\eeq
With the identification $\phi( \mathbf w) = \phi_{1\text{-kink}}^{(0)}(\mathbf w)$, equations (\ref{eq:diffmap}) can be easily integrated yielding
\beqa
\label{eq:map1kink}
z( \mathbf w) &=& w - 4\ta\frac{m}{a\,\beta}\tanh{\[\frac{m}{\beta}\( aw+\frac{1}{a}\bar{w} \)\]} \;, \notag \\ \bar{z}( \mathbf w) &=& \bar{w}- 4\ta\frac{am}{\beta}\tanh{\[\frac{m}{\beta}\( aw+\frac{1}{a}\bar{w} \)\]} \;,
\eeqa
where the constants of integration are fixed consistently with the $\ta=0$ case. 
\begin{figure}[t]
  \centering
  \subfloat[]{\includegraphics[scale=0.20]{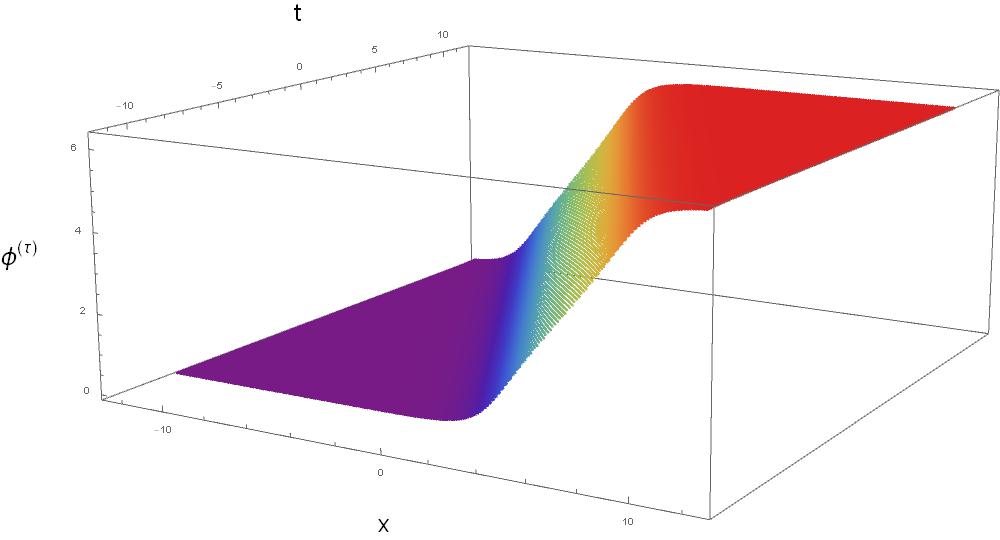}
  \label{fig:1kink_tm025}}
  \hspace{0.5cm}
  \subfloat[]{\includegraphics[scale=0.20]{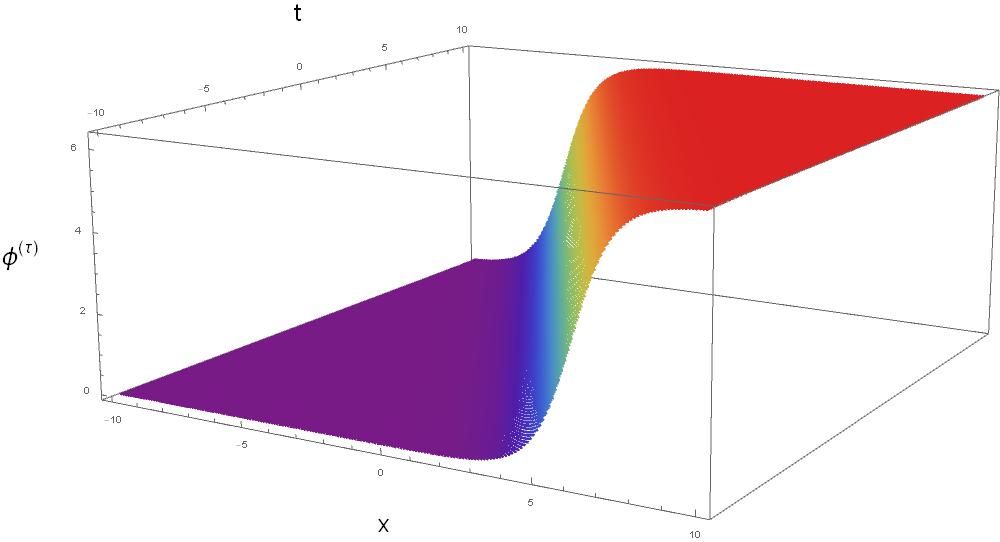}
  \label{fig:1kink_t0}} 
  \vspace{0.5cm}
  \subfloat[]{\includegraphics[scale=0.20]{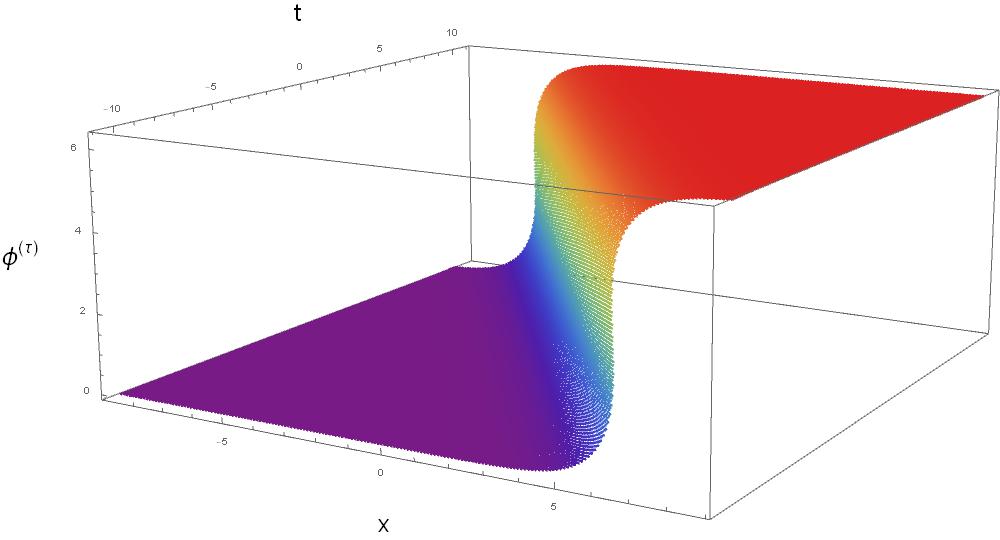}
  \label{fig:1kink_tp0125}}
  \hspace{0.5cm}
  \subfloat[]{\includegraphics[scale=0.20]{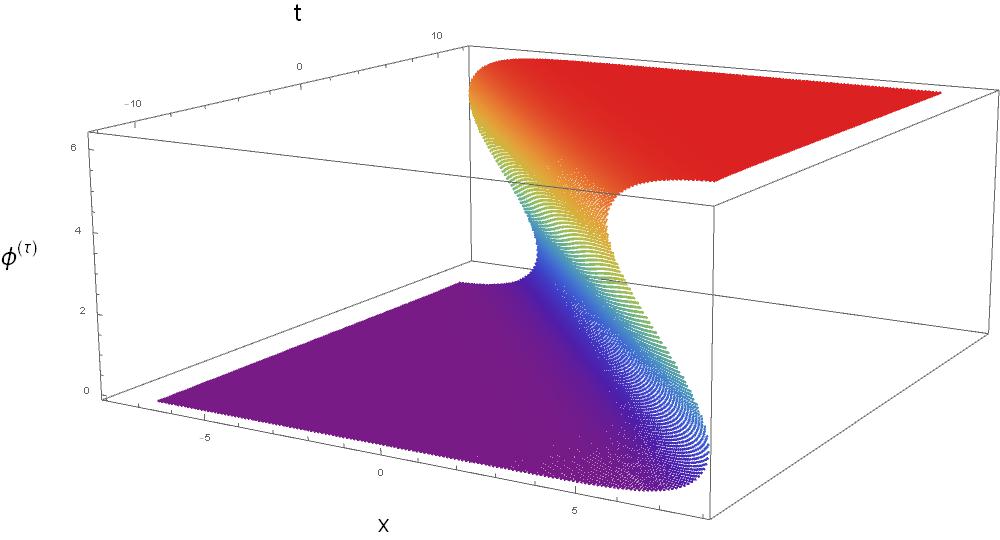}
  \label{fig:1kink_tp03}}
  \vspace{0.1cm}
\caption{The $\textsc{T} \bar{\textsc{T}}$-deformed moving one-kink solution $\(m=\beta=1 \,,\, a=2 \)$, for different values of the perturbation parameter $\ta$. Figure \ref{fig:1kink_t0} represents the undeformed solution. Figure \ref{fig:1kink_tm025} corresponds to $\ta=-1/4$, while Figures \ref{fig:1kink_tp0125} and \ref{fig:1kink_tp03}  correspond to  $\ta=1/8$ and $\ta=1/3$, respectively. Notice that at $\ta=1/8$ a shock-wave singularity occurs.}
  \label{fig:1kink}
\end{figure}
Notice that from (\ref{eq:1kink}) we have
\beq
\frac{m}{\beta}\( aw+\frac{1}{a}\bar{w} \) = \log{\(\tan{\(\frac{\phi_{1\text{-kink}}^{(0)}(\mathbf w)}{4}\)}\)} \;,
\eeq
and thus expressions (\ref{eq:map1kink}) become
\beqa
\label{eq:map1kink2}
z( \mathbf w) = w +4\ta\frac{m}{a\,\beta}\cos{\(\frac{\phi_{1\text{-kink}}^{(0)}( \mathbf w)}{2}\)} \;,\quad
\bar{z}( \mathbf w) = \bar w +4\ta\frac{a\,m}{\beta}\cos{\(\frac{\phi_{1\text{-kink}}^{(0)}( \mathbf w)}{2}\)} \;,
\eeqa
which are easily inverted as
\beqa
\label{eq:map1kinkinv}
w(\mathbf z) &=& z - 4\ta\frac{m}{a\,\beta}\cos{\(\frac{\phi_{1\text{-kink}}^{(0)}\(\mathbf w(\mathbf z) \)}{2}\)} = z - 4\ta\frac{m}{a\,\beta}\cos{\(\frac{\phi_{1\text{-kink}}^{(\tau)}\(\mathbf z\)}{2}\)} \;, \notag \\ 
\bar{w}(\mathbf z) &=& \bar{z} - 4\ta\frac{am}{\beta}\cos{\(\frac{\phi_{1\text{-kink}}^{(0)}\(\mathbf w(\mathbf z) \)}{2}\)} = \bar{z} - 4\ta\frac{am}{\beta}\cos{\(\frac{\phi_{1\text{-kink}}^{(\ta)}\(\mathbf z\)}{2}\)} \;. \notag \\
\eeqa
Finally, plugging (\ref{eq:map1kinkinv}) into (\ref{eq:1kink}) we find 
\beq
\frac{m}{\beta}\( a z+\frac{1}{a}\bar{z} \) = 8\ta\frac{m^2}{\beta^2}\cos{\(\frac{\phi_{1\text{-kink}}^{(\ta)}(\mathbf z)}{2}\)} + \log{\(\tan{\(\frac{\phi_{1\text{-kink}}^{(\ta)}(\mathbf z )}{4}\)}\)} \;,
\eeq
which is exactly the deformed one-kink solution found in \cite{Conti:2018jho}. In Figure \ref{fig:1kink} the  solution is represented for different values of $\ta$. Notice that for negative values of $\ta $ (Figure \ref{fig:1kink_tm025}) the solution stretches w.r.t the undeformed one (Figure \ref{fig:1kink_t0}), while for positive values of $\ta$ (Figures \ref{fig:1kink_tp0125} and \ref{fig:1kink_tp03}) it bends and becomes multi-valued. In particular $\ta=1/8$ (Figure \ref{fig:1kink_tp0125}) is the delimiting value corresponding to a shock wave singularity.
\subsection{The two-kink solution}
Consider now the solution which describes the scattering between two kinks with velocities $v_1$ and $v_2$
\beq
\label{eq:2kink}
\phi_{2\text{-kink}}^{(0)}( \mathbf w) = 4 \arctan\( \frac{a_1+a_2}{a_2-a_1} \frac{e^{\frac{m}{\beta}\(a_1  w +\frac{1}{a_1}\bar w + k_1\)} - e^{\frac{m}{\beta}\(a_2  w +\frac{1}{a_2}\bar w + k_2\)}}{1 + e^{\frac{m}{\beta}\(a_1  w +\frac{1}{a_1}\bar w + k_1\)}e^{\frac{m}{\beta}\(a_2  w +\frac{1}{a_2}\bar w + k_2\)}} \) \;,
\eeq
where again $a_i = \sqrt{\frac{1-v_i}{1+v_i}} \;,\; i=1,2$ , and $k_i \;,\; i=1,2$, are constant phases.
Compared to the one-kink case, this time the sets of differential equations (\ref{eq:diffmap}) are more complicated to integrate. It is useful to parametrize the solutions $z(\mathbf w)$ and $\bar z(\mathbf w)$ of (\ref{eq:diffmap}) in terms of the combinations
\beq
\uZ_i(\mathbf w) = \frac{m}{\beta}\( a_i  w +\frac{1}{a_i}\bar w + k_i \) \;,\; i=1,2 \;.
\eeq
\begin{figure}[t]
  \centering
  \subfloat[]{\includegraphics[scale=0.20]{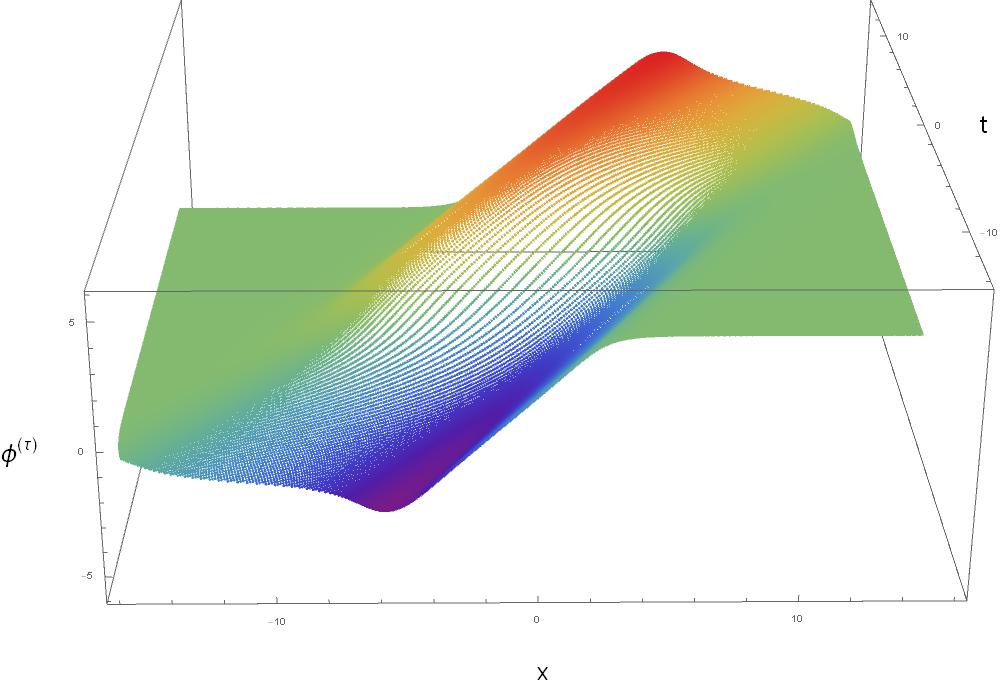}
  \label{fig:2kink_tm025}}
  \hspace{0.5cm}
  \subfloat[]{\includegraphics[scale=0.20]{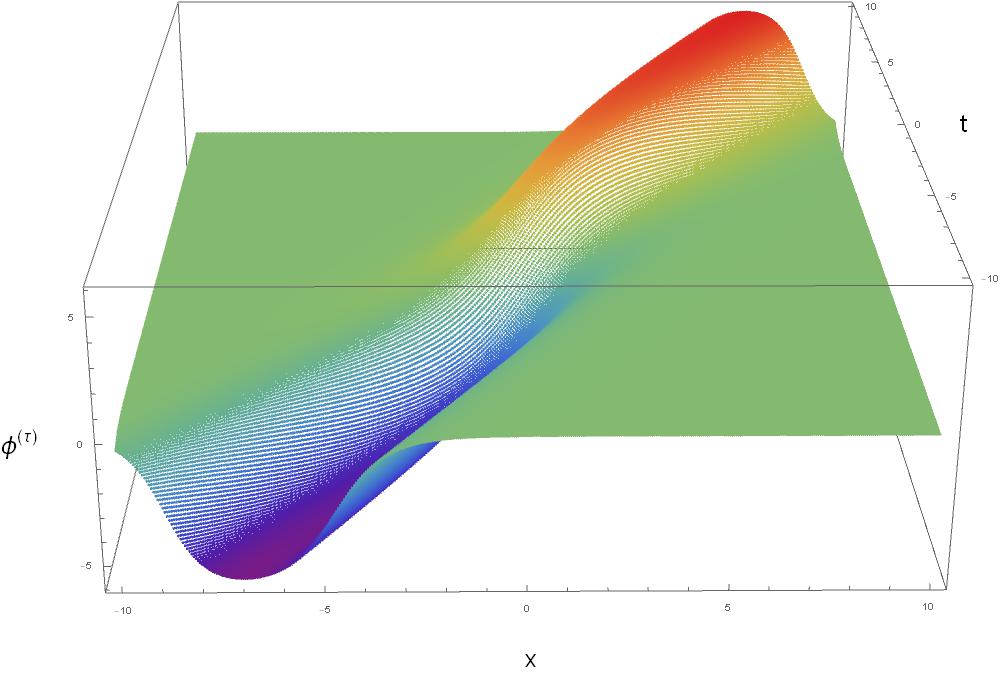}
  \label{fig:2kink_t0}}
  \vspace{0.5cm}
  \subfloat[]{\includegraphics[scale=0.20]{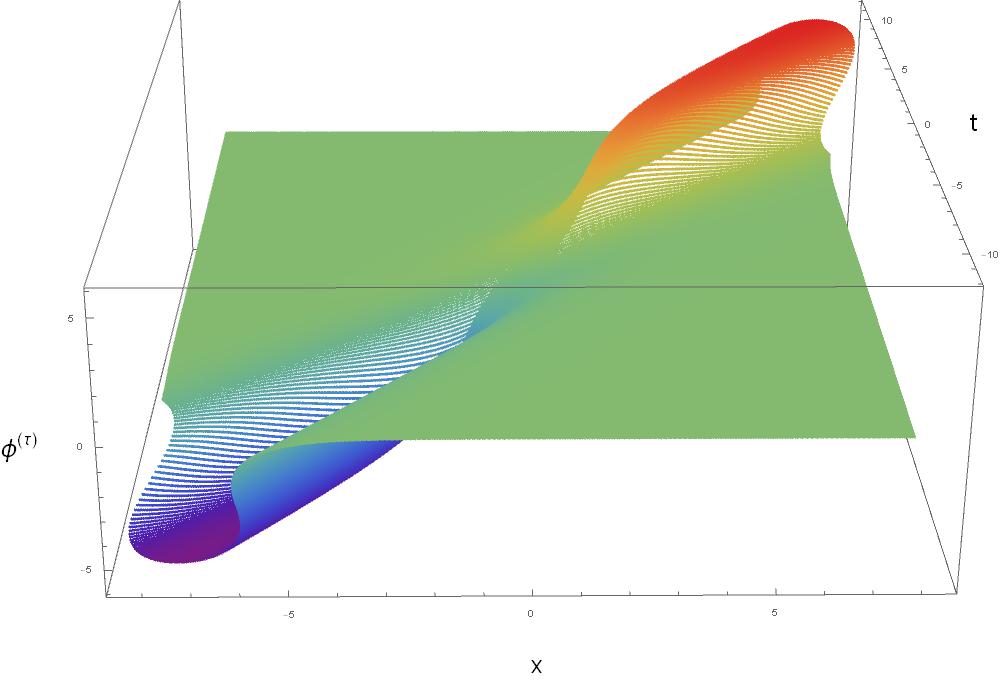}
  \label{fig:2kink_tp01}}
  \hspace{0.5cm}
  \subfloat[]{\includegraphics[scale=0.20]{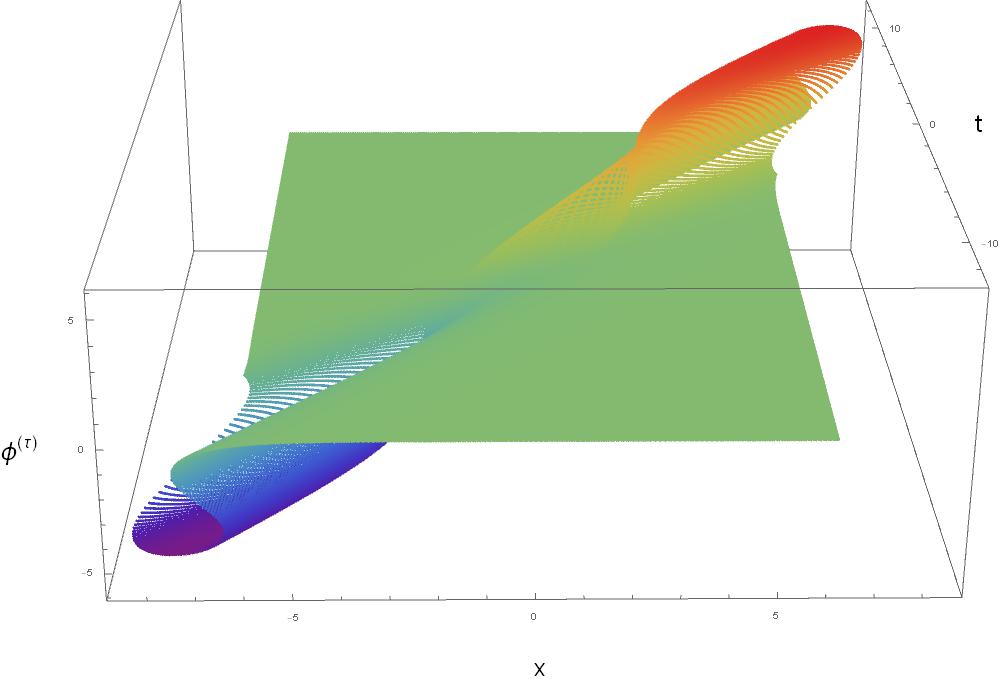}
  \label{fig:2kink_tp016}}
  \vspace{0.1cm}
   \caption{The $\textsc{T} \bar{\textsc{T}}$-deformed two-kink solution $\(m=\beta=1 \,,\, a_1=2 \,,\, a_2=3\)$, for different values of the perturbation parameter $\ta$. 
   Figure \ref{fig:2kink_t0} represents the undeformed solution. Figure \ref{fig:2kink_tm025} corresponds to $\ta=-1/4$, while Figures \ref{fig:2kink_tp01} and  \ref{fig:2kink_tp016} correspond to $\ta$, {\it i.e.} $\ta=1/10$ and $\ta=1/6$, respectively.}
  \label{fig:2kink}
\end{figure}
Performing the change of variables $\mathbf \uZ = (\uZ_1( \mathbf w), \uZ_2(\mathbf w))$
\beqa
\label{eq:wToZi}
\begin{cases}
\frac{\partial z}{\partial \uZ_1}  = \frac{\beta}{m}\frac{a_1\( \frac{\partial z}{\partial w} - a_2^2 \frac{\partial z}{\partial\bar w} \)}{a_1^2-a_2^2} \\
\frac{\partial z}{\partial \uZ_2}  = -\frac{\beta}{m}\frac{a_2\( \frac{\partial z}{\partial w} - a_1^2 \frac{\partial z}{\partial\bar w} \)}{a_1^2-a_2^2}
\end{cases} \;, \quad
\begin{cases}
\frac{\partial\bar z}{\partial \uZ_1} =  \frac{\beta}{m}\frac{a_1\( \frac{\partial\bar z}{\partial w} - a_2^2 \frac{\partial\bar z}{\partial\bar w} \)}{a_1^2-a_2^2} \\
\frac{\partial\bar z}{\partial \uZ_2}  = -\frac{\beta}{m}\frac{a_2\( \frac{\partial\bar z}{\partial w} - a_1^2 \frac{\partial\bar z}{\partial\bar w} \)}{a_1^2-a_2^2}
\end{cases} \;,
\eeqa
and plugging (\ref{eq:diffmap}) into (\ref{eq:wToZi}) with the identification $\phi(\mathbf w ) \equiv \phi_{2\text{-kink}}^{(0)}(\mathbf w)$, we obtain two sets of differential equations which can be solved for $\mathbf z(\mathbf \uZ)$,  giving
\beqa
\label{eq:map2kink}
z(\mathbf \uZ) &=& \frac{\beta}{m}\frac{a_1 \uZ_1-a_2 \uZ_2}{a_1^2-a_2^2} - 4\ta\frac{m}{\beta}\frac{(a_1^2-a_2^2)\(a_1\tanh{\uZ_2}-a_2\tanh{\uZ_1}\)}{a_1a_2\( a_1^2 + a_2^2 -2a_1a_2 \( \sech{\uZ_1}\sech{\uZ_2}+\tanh{\uZ_1}\tanh{\uZ_2} \) \) } \;, \notag \\
\bar{z}(\mathbf \uZ) &=& \frac{\beta}{m}\frac{a_1a_2\(a_1 \uZ_2-a_2 \uZ_1\)}{a_1^2-a_2^2} - 4\ta\frac{m}{\beta}\frac{(a_1^2-a_2^2)\(a_1\tanh{\uZ_1}-a_2\tanh{\uZ_2}\)}{a_1^2 + a_2^2 -2a_1a_2 \( \sech{\uZ_1}\sech{\uZ_2}+\tanh{\uZ_1}\tanh{\uZ_2} \) } \;. \notag \\
\eeqa
As in the previous section, the constants of integration in (\ref{eq:map2kink}) are fixed by imposing the consistency with the $\ta=0$ case. In order to find the deformed two-kink solution $\phi_{2\text{-kink}}^{(\ta)}\( \mathbf z \)=\phi_{2\text{-kink}}^{(0)}\(\mathbf \uZ(\mathbf z)\)$, we should solve (\ref{eq:map2kink}) for $\mathbf \uZ(\mathbf z)$. Since this is analytically very complicated, we resort to numerical inversion. In Figure \ref{fig:2kink} the deformed solution $\phi_{2\text{-kink}}^{(\ta)}\( \mathbf z \)$ is reported for different values of $\ta$. The picture is quite similar to the one-kink case. In fact, for negative values of $\ta$ (Figure \ref{fig:2kink_tm025}) the solution stretches w.r.t. the undeformed one (Figure \ref{fig:2kink_t0}), while for positive values of $\ta$ (Figures \ref{fig:2kink_tp01} and \ref{fig:2kink_tp016}) it bends and again it becomes multi-valued. Unlike the one-kink case, here it is not possible to find analytically the delimiting value of $\ta$ corresponding to the shock singularity.

\subsection{The  breather}
Another interesting solution is the breather with envelope speed $v=0$
\beq
\label{eq:breather}
\phi_{\text{breather}}^{(0)}(\mathbf w) = 4 \arctan\( \tan{\psi}\,\frac{\sin{\(-\frac{m}{\beta}( w-\bar w)\cos{\psi}+\bar k \)}}{\cosh{\(\frac{m}{\beta}( w+\bar w)\sin{\psi}+k \)}} \) \;,
\eeq
where $\psi$ is a parameter related to the period $T$ of one full oscillation via $T = \frac{2\pi}{\cos{\psi}}$ and $k,\bar{k}$ are constant phases. In analogy with the two-kink case, it is useful to use the same strategy and parametrize the solutions $\mathbf z(\mathbf w)$ of (\ref{eq:diffmap}) in terms of
\beq
\uZ(\mathbf w) = \frac{m}{\beta}( w+\bar w)\sin{\psi}+k \;,\quad \bar \uZ(\mathbf w) = -\frac{m}{\beta}( w-\bar w)\cos{\psi}+\bar k \;.
\eeq
Performing the change of variables $\mathbf \uZ( \mathbf w)=(\uZ( \mathbf w), \bar \uZ( \mathbf w))$, one finds
\beqa
\label{eq:wToZ}
\begin{cases}
\frac{\partial z}{\partial \uZ}= \frac{\beta}{m}\frac{1}{2\sin{\psi}}\( \frac{\partial z}{\partial w} + \frac{\partial z}{\partial\bar w} \) \\
\frac{\partial z}{\partial \bar \uZ}  = \frac{\beta}{m}\frac{1}{2\cos{\psi}}\( -\frac{\partial z}{\partial w} + \frac{\partial z}{\partial\bar w} \)
\end{cases} \;, \quad
\begin{cases}
\frac{\partial\bar z}{\partial \uZ}  = \frac{\beta}{m}\frac{1}{2\sin{\psi}}\( \frac{\partial \bar z}{\partial w} + \frac{\partial \bar z}{\partial\bar w} \) \\
\frac{\partial\bar z}{\partial \bar \uZ}  = \frac{\beta}{m}\frac{1}{2\cos{\psi}}\( -\frac{\partial \bar z}{\partial w} + \frac{\partial \bar z}{\partial\bar w} \)
\end{cases} \;,
\eeqa
and again plugging (\ref{eq:diffmap}) into (\ref{eq:wToZ}) with the identification $\phi(\mathbf w )\equiv\phi_{\text{breather}}^{(0)}(\mathbf w)$,
\begin{figure}[t]
 \centering
  \subfloat[]{\includegraphics[scale=0.20]{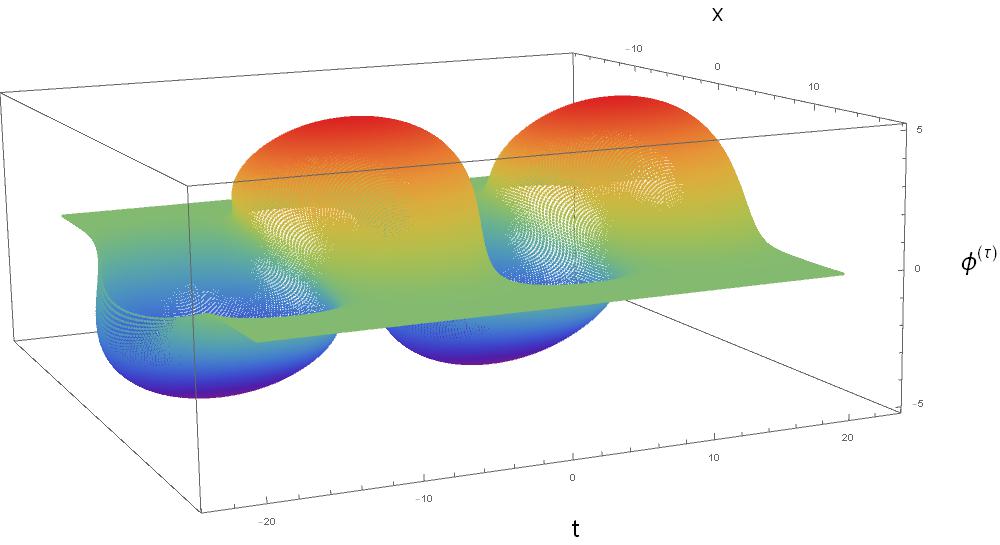}
  \label{fig:breather_tm05}}
  \hspace{0.5cm}
  \subfloat[]{\includegraphics[scale=0.20]{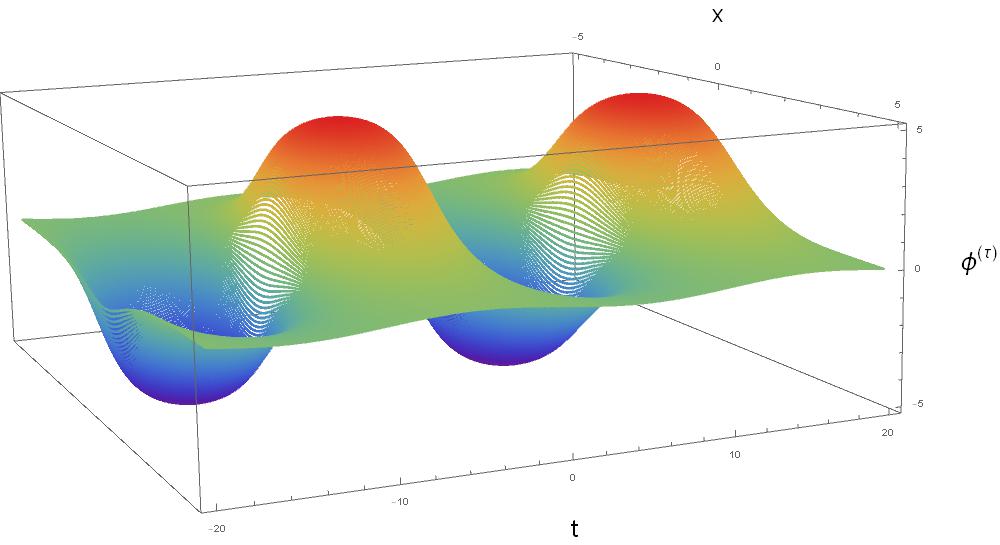}
  \label{fig:breather_t0}}
  \vspace{0.5cm}
  \subfloat[]{\includegraphics[scale=0.20]{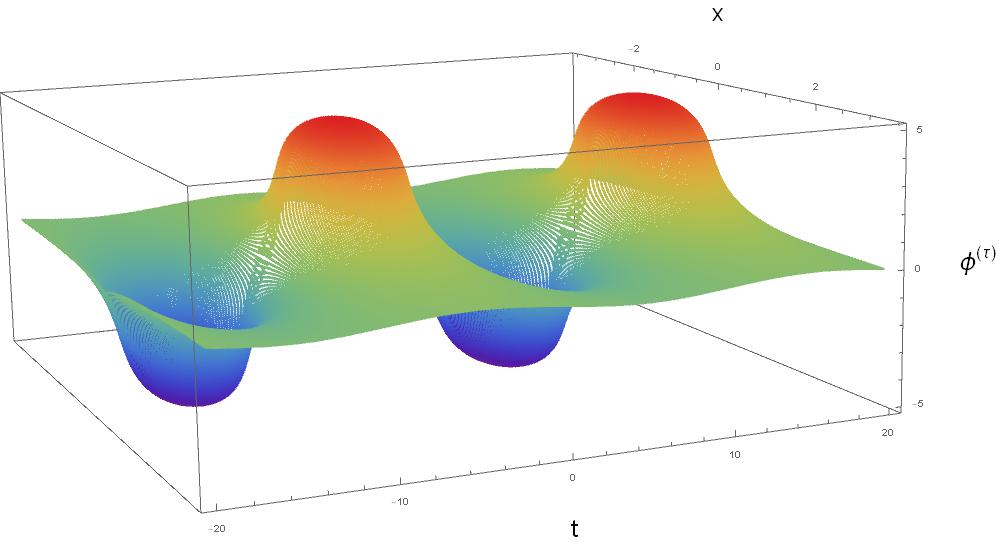}
  \label{fig:breather_tp01}}
  \hspace{0.5cm}
  \subfloat[]{\includegraphics[scale=0.20]{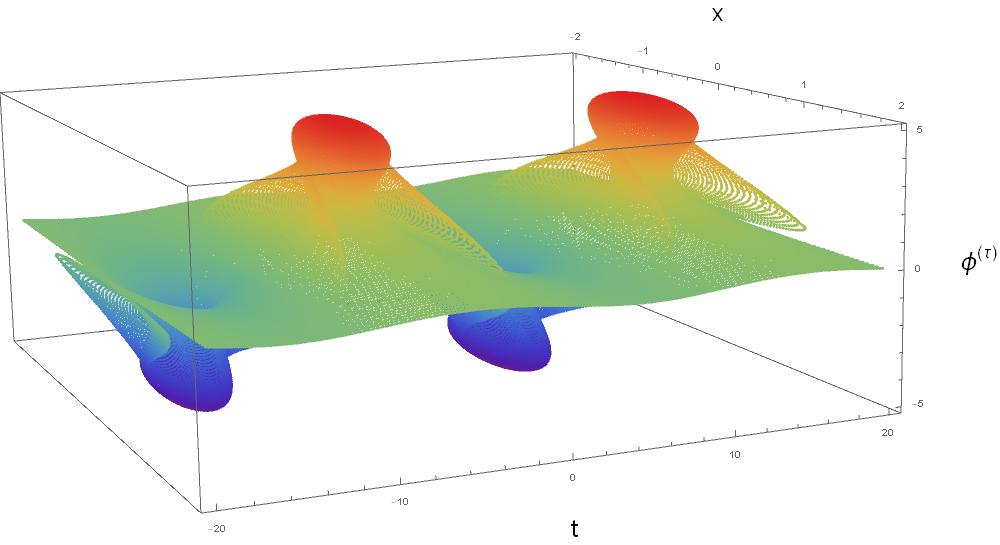}
  \label{fig:breather_tp02}}
  \vspace{0.1cm}
   \caption{The $\textsc{T} \bar{\textsc{T}}$-deformed stationary  breather solution with envelope speed $v=0$ ($m=\beta=1 \,,\, \psi=\frac{2}{5}\pi$), for different values of the perturbation parameter $\ta$. Figure \ref{fig:breather_t0} represents the undeformed solution,  Figure \ref{fig:breather_tm05} corresponds to $\ta=-1/2$, while Figures \ref{fig:breather_tp01} and \ref{fig:breather_tp02} correspond to $\ta=1/10$ and $\ta=1/5$, respectively.}
  \label{fig:breather}
\end{figure}
one gets two sets of differential equations which can be solved for $\mathbf z(\mathbf \uZ)$ giving
\beqa
\label{eq:mapbreather}
z(\mathbf \uZ) &=& \frac{\beta}{m}\(\frac{\uZ}{2\sin{\psi}} - \frac{\bar \uZ}{2\cos{\psi}}\) -8\ta \frac{m}{\beta}\sin{\psi}\frac{\cos{\bar \uZ}}{\cosh{\uZ}}\frac{\sec{\bar \uZ}\sinh{\uZ}+\sech{\uZ}\sin{\bar \uZ}\tan{\psi}}{1+\(\tan{\psi}\sin{\bar \uZ}\sech{\uZ}\)^2} \notag \;, \\
\bar z(\mathbf \uZ) &=& \frac{\beta}{m}\(\frac{\uZ}{2\sin{\psi}} + \frac{\bar \uZ}{2\cos{\psi}}\) -8\ta \frac{m}{\beta}\sin{\psi}\frac{\cos{\bar \uZ}}{\cosh{\uZ}}\frac{\sec{\bar \uZ}\sinh{\uZ}-\sech{\uZ}\sin{\bar \uZ}\tan{\psi}}{1+\(\tan{\psi}\sin{\bar \uZ}\sech{\uZ}\)^2} \notag \;. \\
\eeqa
As for the two-kink example, the constants of integration in (\ref{eq:mapbreather}) are fixed according to the $\ta=0$ case, and again the solution $\mathbf \uZ(\mathbf z)$ to (\ref{eq:mapbreather}) is computed numerically. The deformed solution $\phi_{\text{breather}}^{(\ta)}(\bf z)$ is displayed in Figure \ref{fig:breather} for different values of $\ta$. The result is similar to the previous cases: the solution stretches for negative values of $\ta$ (Figure \ref{fig:breather_tm05}) and it bends for positive values of $\ta$ (Figure \ref{fig:breather_tp01} and \ref{fig:breather_tp02}) w.r.t. the undeformed one (Figure \ref{fig:breather_t0}). However, notice that in this case the shock phenomenon occurs in both positive and negative directions of $\ta$, and consequently the solution becomes multi-valued (Figures \ref{fig:breather_tm05} and \ref{fig:breather_tp02}) for $|\ta|$ sufficiently large.
\section{The shock-wave phenomenon and the Hagedorn-type transition }
\label{sec:Hagedorn}
In this section we will discuss the emergence of critical phenomena in the classical solutions, \textit{i.e.} the shock-wave singularity and the square root-type transition, and comment on the relations among them. We will use as a guide example the stationary $\TbT$-deformed elliptic solution of the sG model derived in \cite{Conti:2018jho}, where we set $\rho=1/\kappa>0$ and $m=\beta=1$,
\beq
\label{eq:stair}
x = \frac{1}{2\sqrt{\rho}}\[\(\rho+4\ta\)\, \mathbf{F}\left(\frac{\phi(x)}{2}\,\vline-\rho\right)-8\tau\, \mathbf{E}\left(\frac{\phi(x)}{2}\,\vline-\rho\right)\] \;,
\eeq
defined on a cylinder of radius $R$ fixed. Due to the following properties of the elliptic functions
\beqa
\mathbf{F}(z+n\pi\,|\gamma) &=& \mathbf{F}(z\,|\gamma) + 2n\,\mathbf{K}(\gamma) \;,\notag \\ \mathbf{E}(z+n\pi\,|\gamma) &=& \mathbf{E}(z\,|\gamma) + 2n\,\mathbf{E}(\gamma) \;,\quad z,\gamma\in\mathbb{C} \;,\quad n\in\mathbb{Z} \;,
\eeqa
the solution $\phi(x)$ can be interpreted as a stationary 1-kink with twisted boundary conditions
\beq 
\phi(x+R) = \phi(x) + 2\pi \;,
\eeq
where the radius $R$ is
\beq
\label{eq:R}
R = \frac{1}{\sqrt{\rho}}\,\bigl(\(\rho + 4\ta\)\,\mathbf{K}\left(-\rho\right) -8\ta\,\mathbf{E}\left(-\rho\right)\bigr) \;.
\eeq
We stress that $R$ is kept fixed while $\rho=\rho(\ta,R)$ is considered as a function of $\ta$ and $R$, defined implicitly through (\ref{eq:R}). Differentiating both sides of (\ref{eq:R}) w.r.t. $\tau$ and $R$ and solving for $\partial_\ta\rho$ and $\partial_R\rho$ one finds
\beq
\label{eq:ktauT}
\partial_\ta \rho = -\frac{8 \rho\,(1+\rho) \bigl(2 \mathbf{E}\left(-\rho\right)-\mathbf{K}\left(-\rho\right)\bigr)}{(\rho + 4\ta)\, \mathbf{E}\left(-\rho\right)} \;,\quad
\partial_R \rho = \frac{2 \rho^{3/2}\,(1 + \rho)}{(\rho + 4 \ta)\,\mathbf{E}\left(-\rho\right)} \;.
\eeq
We shall now compute the energy on the cylinder. The components of the Hilbert stress-energy tensor $T_{\mu\nu}^{(\ta)}$ are
\beqa
\label{eq:H}
T_{22}^{(\ta)} &\equiv& \mathcal{H}^{(\ta)} = \frac{V}{1-\ta V} + \frac{1 + \ta(1-\ta V)\,\phi_x^2 - S}{2S\,\ta(1-\ta V)} = \frac{2\,(2+\rho V)}{\rho\,(1-2\tau V) - 4\ta} \;, \\
\label{eq:calP}
T_{12}^{(\ta)} &=& T_{21}^{(\ta)} \equiv \mathcal{P}^{(\ta)} = -\frac{\phi_t\,\phi_x}{2S} = 0 \;, \\
\label{eq:T11}
T_{11}^{(\ta)} &=& -\frac{V}{1-\ta V}-\frac{1-\ta(1-\ta V)\,\phi_t^2-S}{2S\,\ta(1-\ta V)} = \frac{4}{\rho+4\ta} \;,
\eeqa
where we used the following expressions for $\phi_t$ and $\phi_x$ derived from (\ref{eq:stair})
\beq
\label{eq:phixt}
\phi_t = 0 \;,\quad \phi_x = \frac{2\sqrt{\rho}\,\sqrt{4 + \rho V}}{\rho\,(1-2\ta V) - 4 \ta} \;,
\eeq
and
\beq 
S = \sqrt{1+\ta(1-\ta V)\( \phi_x^2 - \phi_t^2 \)} = \frac{\rho+4\ta}{\rho\,(1-2\ta V)-4\ta} \;.
\eeq
Notice that the apparent pole singularity at $\ta = 1/V$ in $T_{11}^{(\ta)}$ and $T_{22}^{(\ta)}$ disappears once (\ref{eq:phixt}) is used in (\ref{eq:H}) and (\ref{eq:T11}). Finally the energy and momentum at finite volume $R$ are
\beqa
\label{eq:E}
E^{(\ta)} &=& \int_{x_0}^{x_0 + R} \mathcal{H}^{(\ta)}(x)\,dx =  \int_{\phi(x_0)=0}^{\phi(x_0+R)=2\pi} \frac{\mathcal{H}^{(\ta)}(\phi)}{\phi_x}\,d\phi = 
\frac{4}{\sqrt{\rho}}\,\bigl( 2\,\mathbf{E}(-\rho) - \mathbf{K}(-\rho) \bigr) \;, \\
\label{eq:P}
P^{(\ta)} &=& \int_{x_0}^{x_0 + R} \mathcal{P}^{(\ta)}(x)\,dx =  \int_{\phi(x_0)=0}^{\phi(x_0+R)=2\pi} \frac{\mathcal{P}^{(\ta)}(\phi)}{\phi_x}\,d\phi = 0 \;, \\
\label{eq:K}
K^{(\ta)} &=& \int_{x_0}^{x_0 + R} T_{11}^{(\ta)}(x)\,dx = \int_{\phi(x_0)=0}^{\phi(x_0+R)=2\pi} \frac{T_{11}^{(\ta)}(\phi)}{\phi_x}\,d\phi = \frac{4R}{\rho+4\ta} \;,
\eeqa
where $x_0 = 0\,(\text{mod}\,R)$. From (\ref{eq:ktauT}), (\ref{eq:E}) and (\ref{eq:K}) one can prove that the energy fulfils the Burgers equation (\ref{eq:Burgers}) with $P_n=0$
\beq
\label{eq:EPBurgers}
\partial_\ta E^{(\ta)} = \frac{1}{2}E^{(\ta)}\partial_R E^{(\ta)} = -\frac{1}{R}\,\det\(\int_{x_0}^{x_0+R} T_{\mu\nu}^{(\ta)}(x) \,dx \) = -\int_{x_0}^{x_0+R} \det\( T_{\mu\nu}^{(\ta)}(x) \)dx \;,
\eeq
where the last equality in (\ref{eq:EPBurgers}) shows the factorization property of the $\TbT$ operator at the classical level. Since the energy $E^{(\ta)}$ fulfils a Burgers equation, it is expected to have a square root-type singularity.\footnote{It is worth to notice that the unperturbed energy $E^{(0)}$ displays the following divergent behavior for small $R$
\beq
E^{(0)} = \frac{\pi^2}{R} + 2R - \frac{R^3}{2\pi^2} + \mathcal O\left(R^7\right)\;,
\eeq
which resembles that of a CFT.} The critical radius $R_c(\ta)$ corresponds to a value of $R$ such that the first derivative of $E^{(\ta)}(R)$ w.r.t. $R$ diverges. One easily checks that
\beq
\partial_R E^{(\ta)} = -\frac{4}{\rho+4\ta}\;,
\eeq
thus the first derivative is divergent at the radius $R_c(\tau)$ defined through the equation
\beq
\label{eq:Hagcond}
\rho\bigl(\tau,R_c(\tau)\bigr) = -4\tau \;.
\eeq
According to (\ref{eq:R}) and (\ref{eq:E}), the critical radius and the corresponding energy turn out to be
\beq
\label{eq:EH}
R_c(\ta) = 4\sqrt{-\ta}\,\mathbf{E}\left(4\ta\right) \;,\quad E^{(\ta)}_c\equiv E^{(\ta)}(R_c) = 
\frac{2}{\sqrt{-\ta}}\,\bigl( \mathbf{K}(4\ta) -2\,\mathbf{E}(4\ta) \bigr) \;.
\eeq
To find the behavior of $E^{(\ta)}$ as a function of $R$ close to the branch singularity $R_c$, we first expand $R$ and $E^{(\ta)}$ in powers of the small quantity $\varepsilon=\rho+4\ta$
\beqa
\label{eq:RHEH}
R-R_c &=& \frac{R_c}{128\ta^2\,(1-4\ta)}\,\varepsilon^2 + \mathcal{O}(\varepsilon^3) \;, \notag \\
E^{(\ta)} - E^{(\ta)}_{c} &=&  \frac{R_c}{16\ta^2\,(1-4\ta)}\,\varepsilon + \mathcal{O}(\varepsilon^2) \;,
\eeqa
then, removing $\varepsilon$, one finds
\beq
\label{eq:EHfin}
E^{(\tau)} - E^{(\ta)}_{c} = \pm\frac{\sqrt{R_c}}{\tau \sqrt{2-8\tau}}\,\sqrt{R-R_c} + \mathcal O\left(R-R_c\right)\;,
\eeq
which gives a square root branch point at $R_c$ for the energy.\\
\begin{figure}[t]
  \centering
  \subfloat[]{\includegraphics[scale=0.60]{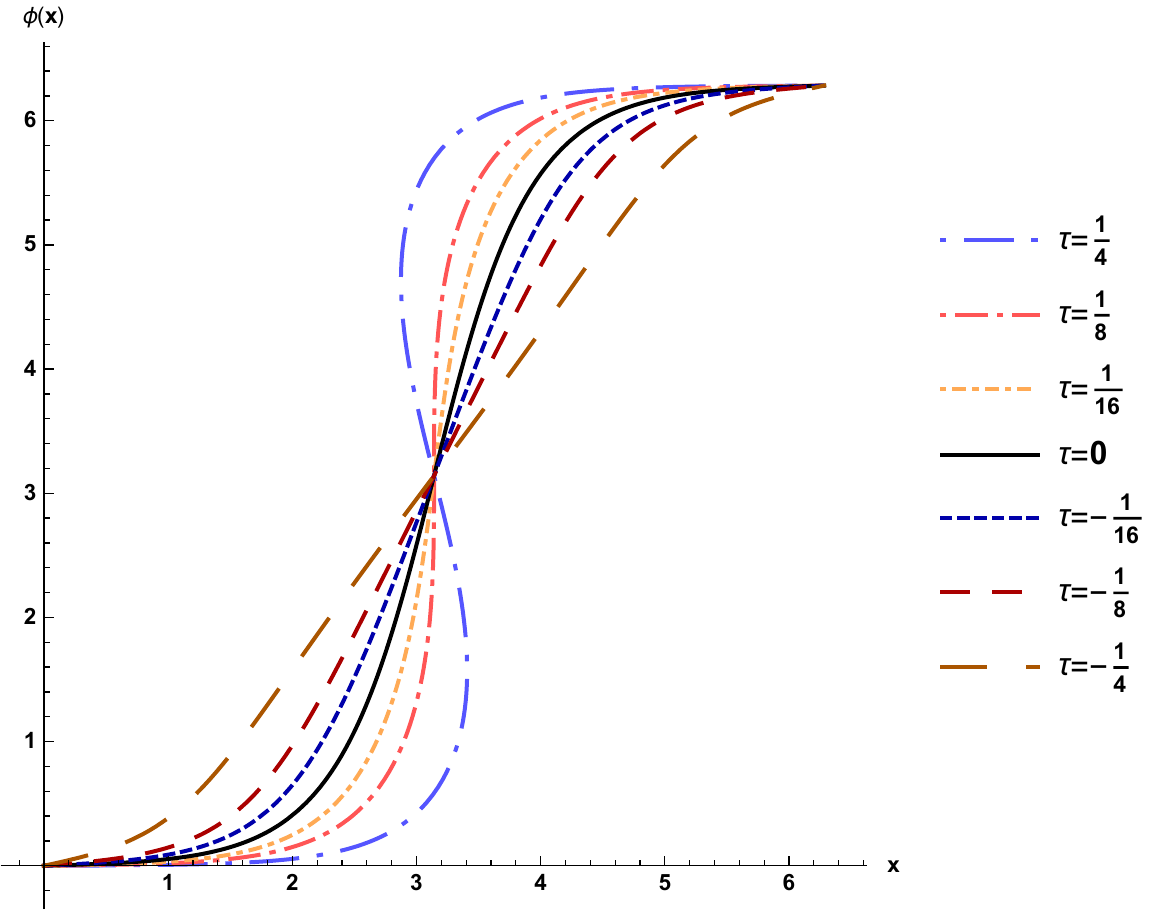}
  \label{fig:sg_kink}}
  \hspace{0.5cm}
  \subfloat[]{\includegraphics[scale=0.60]{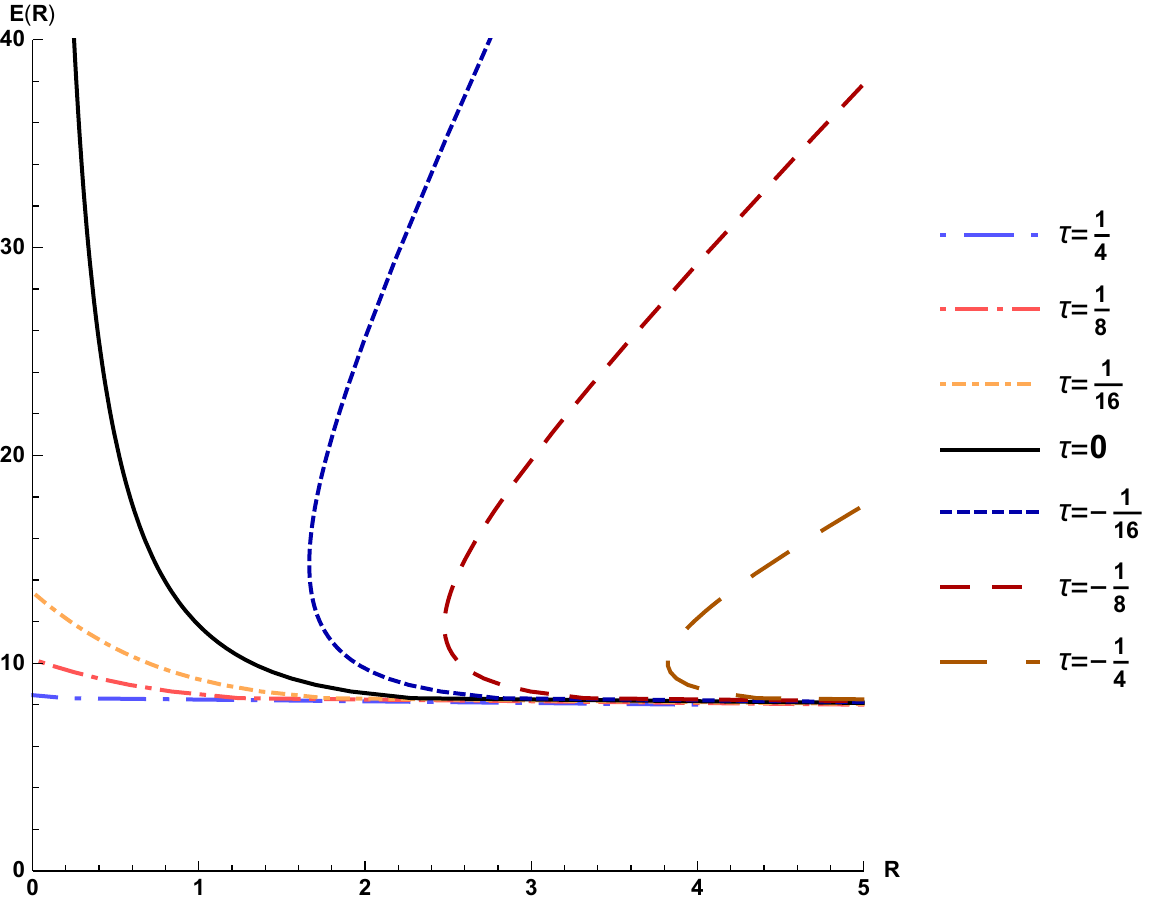}
  \label{fig:sg_kink_energy}}
  \vspace{0.1cm}
\caption{The kink solution to the $\TbT$-deformed sG model on a cylinder of radius $R$ (a) and the corresponding energies as functions of $R$ (b).}
  \label{fig:1kinks}
\end{figure}
Now we would like to briefly discuss the effect of the shock-wave singularities of the deformed solution on the Hamiltonian density. To compute the range of values of $\ta$ where the solution becomes multi-valued, we first identify the zeros of $\text{Det}\(\mathcal{J}^{-1}\)$: 
\beq 
\label{eq:shockcond}
\text{Det}\(\mathcal{J}^{-1}\) = 0 \quad\Longleftrightarrow\quad x = \frac{\sqrt{\rho}}{2}\,\text{dn}^{-1}\left(\pm\sqrt{\frac{\rho + 4\ta}{8\ta}}\;\vline - \rho\right) \;,
\eeq
where $\text{dn}^{-1}(z\;|\gamma)$ is the inverse of the Jacobi elliptic function $\text{dn}(z\;|\gamma)$. From the reality properties of $\text{dn}^{-1}(z\;|\gamma)$ it follows that $x$ is real for
\beqa
\label{eq:rangetau}
\rho > 0 \quad\wedge\quad \ta_1^* = \frac{\rho}{4+8\rho} < \ta < \frac{\rho}{4} = \ta_2^* \;,
\eeqa
where the critical values\footnote{Notice that, in the $\rho\to\infty$ limit, one recovers the $1$-kink solution, and the critical range reduce to $\ta>\ta_1^*=\frac{1}{8} \;,$
since $\ta_2^*\to\infty$.} $\ta_1^*$ and $\ta_2^*$ corresponds to shock-wave singularities of the solution at $\phi = \pi$ and $\phi = 0,2\pi$, respectively. The Hamiltonian density (\ref{eq:H}) is indeed singular when
\beq 
\label{eq:HPpole}
\ta = \frac{\rho}{4 + 2\rho V} = \frac{\rho}{4 + 8\rho\,\sin^2{(\phi/2)}} \;,
\eeq
which corresponds to the range of singular values of $\ta$ (\ref{eq:rangetau}) as $\phi$ interpolates from $0$ to $2\pi$. However, it is important to stress that these branching singularities do not affect the total energy (\ref{eq:E}), which remains smooth in $\ta$, since the singularities cancel out when dividing by $\phi_x$ in (\ref{eq:E}). In Figure \ref{fig:1kinks} we displayed the behaviour of $\phi(x)$ (Figure \ref{fig:sg_kink}) and $E^{(\ta)}(R)$ (Figure  \ref{fig:sg_kink_energy}) for various values of $\ta$. We see that the shock-wave phenomenon and the square root-type singularity occur at positive and negative values of $\ta$, respectively.

\section{Conclusions}
Starting from the $\TbT$-deformed Lagrangians proposed in \cite{Cavaglia:2016oda,Conti:2018jho,Bonelli:2018kik}, the main result of this article is the direct derivation of the exact one-to-one map between solutions of the unperturbed and deformed equations of motion, which takes the general form  (\ref{eq:map1},\ref{eq:solutiontrans}). The result matches the topological gravity  predictions of  \cite{Dubovsky:2017cnj, Dubovsky:2018bmo} but it  should be possible to obtain the fundamental equations (\ref{eq:map1},\ref{eq:solutiontrans},\ref{eq:metric1}) also by working within the framework introduced by Cardy in \cite{Cardy:2018sdv}.

We initially arrived to this  conclusion by studying the well known classical relation between sine-Gordon, the  associated Lax operators and pseudo-spherical surfaces embedded in $\RR^3$.
We think that this alternative and more explicit approach to the problem  may provide a complementary point of view compared to \cite{Dubovsky:2017cnj,Dubovsky:2018bmo} and open the  way to the implementation of further integrable model tools, such as the Inverse Scattering Method and the ODE/IM  correspondence within the $\TbT$/JT framework.   

There are many theoretical aspects that deserve to be further explored. First of all, it would be conceptually very  important to study  fermionic theories and supersymmetric sigma models.
In \cite{Bonelli:2018kik},  it was argued that for the $\TbT$-perturbed Thirring model the Lagrangian truncates at second order in $\ta$, such a truncation is not totally surprising, however  the sine-Gordon Lagrangian is instead deformed in an highly non trivial way and it would be nice to identify the mechanism which allows to preserve the quantum equivalence between the two systems. Secondly, it would important to continue the investigation  of deformed 2D Yang-Mills \cite{Conti:2018jho}, along the lines started  in the interesting recent work \cite{Santilli:2018xux}. These studies might also serve as a guide for the  inclusion of the $\TbT$  inside the Wilson Loop/Scattering Amplitude setup \cite{Alday:2009dv,Alday:2010vh} (see also the  remarks in the outlook section of  \cite{Santilli:2018xux}).

Finally,  it would also be interesting to study the generalisation of our results to the $\JbT$ case described in \cite{Bzowski:2018pcy, Guica:2017lia,Chakraborty:2018vja,Apolo:2018qpq, Aharony:2018ics} and to check whether for any of the  higher-dimensional  models  discussed in \cite{Cardy:2018sdv, Bonelli:2018kik, Taylor:2018xcy, Conti:2018jho} there could exist a map, between deformed and undeformed solutions,  similar to  equations  (\ref{eq:map1},\ref{eq:solutiontrans}).

\medskip
\noindent{\bf Note:} We have recently been informed that the coordinate map 
between deformed and undeformed classical Lagrangian systems  was also independently introduced by Chih-Kai Chang and studied  in an on-going research project  involving also  Christian Ferko and Savdeep Sethi.

\medskip
\noindent{\bf Acknowledgments --}
We are especially grateful to  Sasha Zamolodchikov, Sylvain Lacroix for inspiring discussions and help, and to Sergei Dubovsky for kindly  guiding us  through 
the recent literature connecting the $\TbT$ perturbation to JT gravity. We also thank Andrea Cavagli\`a, Riccardo Borsato, Chih-Kai Chang, Yunfeng Jiang, Zohar Komargodski, Marc Magro,  Marc Mezei, Leonardo Santilli, Alessandro Sfondrini, Istv\'an Sz\'ecs\'enyi and Miguel Tierz for useful discussions on related topics.
This project was partially supported by the INFN project SFT, the EU network GATIS+, NSF Award  PHY-1620628, and by the FCT Project PTDC/MAT-PUR/30234/2017 ``Irregular connections on algebraic curves and Quantum Field Theory".
%

%\appendix
\appendix
\section{Short review on surfaces embedded in $\mathbb R^3$}
The purpose of this appendix is to briefly review the basic concepts related to the classical theory of surfaces embedded in the Euclidean space $\mathbb R^3$. We will follow the standard constructive approach which can be found, for example, in \cite{rogers2002backlund}. Let us start by considering a surface $\Sigma$ together with the vector-valued function $\mathbf r\left(\mathbf z\right)\in\mathbb R^3$, describing its embedding into $3$-dimensional flat space. It is clear that the two vectors
\begin{equation}
    \mathbf r_{\mu} = \frac{\partial}{\partial z^{\mu}}\mathbf r\;,\quad \mu = 1,2\;,
\end{equation}
span the tangent plane $T_P\Sigma$ to the surface at any non-critical point $P\in\Sigma$.\footnote{A critical point of a surface is, in this context, defined as a point $P_c$ such that $\mathbf r_1\left(\mathbf z_c\right) = \mathbf r_2\left(\mathbf z_c\right)$.} We will disregard the subtleties arising with the presence of critical points and suppose that $\mathbf r_1\left(\mathbf z\right) \neq \mathbf r_2\left( \mathbf z \right)$ for all points $P\in\Sigma$. This basis of $T_P\Sigma$ can be improved to a basis $\sigma$ of $\mathbb R^3$ by adding the unit normal vector $\mathbf n$
\begin{equation}
    \sigma = \left\lbrace
        \mathbf r_1 , \mathbf r_2 , \mathbf n
    \right\rbrace\;,\qquad \mathbf n = \frac{\mathbf r_1 \times \mathbf r_2}{\left\vert \mathbf r_1 \times \mathbf r_2 \right\vert}\;.
    \label{eq:basisR3}
\end{equation}
The surface $\Sigma$ inherits a metric structure from the ambient space $\mathbb R^3$ and its line element, also known as \emph{first fundamental quadratic form}, is
\begin{equation}
    \textsc I \equiv ds^2 = d\mathbf r \cdot d\mathbf r = \mathbf r_{\mu} \cdot \mathbf r_{\nu} dz^{\mu} dz^{\nu}\;.
    \label{eq:I}
\end{equation}
The tensor
\begin{equation}
    g_{\mu\nu} = \mathbf r_{\mu}\cdot \mathbf r_{\nu}\;,
\label{eq:metric_tensor}
\end{equation}
is called \emph{first fundamental tensor} or \emph{metric tensor} of the surface $\Sigma$. According to the classical theorem by Bonnet \cite{Bonn_867} any surface embedded in flat $3$-space is uniquely determined, up to isometries, by the first and the \emph{second fundamental quadratic form}, defined as
\begin{equation}
    \textsc{II} = - d\mathbf r\cdot d\mathbf n = - \mathbf r_{\nu} \cdot \left(\frac{\partial}{\partial z^{\mu}} \mathbf n\right) \,dz^{\mu}dz^{\nu} = \left(\frac{\partial}{\partial z^{\mu}} \mathbf r_{\nu}\right) \cdot \mathbf n \,dz^{\mu}dz^{\nu}\;.
    \label{eq:II}
\end{equation}
The tensor
\begin{equation}
    d_{\mu\nu} = \left(\frac{\partial}{\partial z^{\mu}} \mathbf r_{\nu}\right) \cdot \mathbf n\;,
\label{eq:second_fund_tensor}
\end{equation}
describes the projection of the vectors $\frac{\partial}{\partial z^{\mu}} \mathbf r_{\nu}\left(P\right)$ on the normal direction and tells us how much the surface curves away from the tangent space in an infinitesimal interval around the point $P$. These two tensors can be combined into the object
\begin{equation}
    s_{\mu}^{\nu} = d_{\mu\rho}g^{\rho\nu}\;,\qquad g_{\mu\rho}g^{\rho\nu} = \delta_{\mu}^{\nu}\;,
\end{equation}
known as \emph{shape} or \emph{Weingarten operator}, whose eigenvalues $\kappa_1,\,\kappa_2$ are the \emph{principal curvatures} of the surface $\Sigma$. The latter  quantities are geometric invariants, meaning that they do not change under coordinate transformations. Usually they are combined into the \emph{Gauss} and \emph{mean} curvatures
\begin{equation}
    K = \kappa_1\kappa_2 = \textrm{Det}\(s_{\mu}^{\nu}\)\;,\quad H = \frac{\kappa_1 + \kappa_2}{2} = \frac{1}{2} s_{\mu}^{\mu}\;.
\label{eq:Gauss_mean_curv_def}
\end{equation}

The tensors $g_{\mu\nu}$ and $d_{\mu\nu}$ determine the structural equations for embedded surfaces, comprising the \emph{Gauss equations}
\begin{equation}
    \frac{\partial}{\partial z^{\mu}} \mathbf r_{\nu} = \Gamma_{\mu\nu}^{\rho} \mathbf r_{\rho} + d_{\mu\nu} \mathbf n\;,
\label{eq:Gauss_eq_GWsystem}
\end{equation}
and the \emph{Weingarten equations}
\begin{equation}
    \frac{\partial}{\partial z^{\mu}} \mathbf n = s_{\mu}^{\nu} \mathbf r_{\nu}\;,
\label{eq:Weingarten_eq_GWsystem}
\end{equation}
where we introduced the \emph{Christoffel symbols} for the metric
\begin{equation}
    \Gamma_{\mu\nu}^{\rho} = \frac{1}{2}g^{\rho\sigma}\left(\frac{\partial}{\partial z^{\nu}} g_{\mu\sigma} + \frac{\partial}{\partial z^{\mu}} g_{\nu\sigma} - \frac{\partial}{\partial z^{\sigma}} g_{\mu\nu}\right)\;.
\end{equation}
These equations describe how the frame $\sigma$ moves on the surface and can be collected into the following linear system
\begin{equation}
    \frac{\partial}{\partial z^{\mu}} \sigma = U_{\mu} \sigma\;,
\end{equation}
with\footnote{Note that $\Gamma_{\mu\nu}^{\rho} = \Gamma_{\nu\mu}^{\rho}$ and $d_{\mu\nu} = d_{\nu\mu}$.}
\begin{equation}
    U_1 = \left(\begin{array}{c c c}
        \Gamma_{11}^1 & \Gamma_{11}^2 & d_{11} \\
        \Gamma_{12}^1 & \Gamma_{12}^2 & d_{12} \\
        -s_1^1 & -s_1^2 & 0
    \end{array}\right)\;,\qquad U_2 = \left(\begin{array}{c c c}
        \Gamma_{12}^1 & \Gamma_{12}^2 & d_{12} \\
        \Gamma_{22}^1 & \Gamma_{22}^2 & d_{22} \\
        -s_2^1 & -s_2^2 & 0
    \end{array}\right)\;.
\end{equation}
These structural equations are subject to a set of compatibility conditions called \emph{Gauss-Mainardi-Codazzi} (GMC) system, which takes the form of a zero curvature condition on the matrices $U_{\mu}$
\begin{equation}
    \partial_2 U_1 - \partial_1 U_2 +\left[U_1,U_2\right] = 0\;.
\label{eq:GMC}
\end{equation}
Note that the matrices $U_\mu$ do not form a Lax pair in the usual sense, since no spectral parameter is present. Moreover, these matrices do not belong to any particular semi-simple Lie algebra. Specialising this general construction to the sine-Gordon case, we will show how to build a proper Lax pair out of the matrices $U_\mu$.\\
As a first example, consider a pseudo-spherical surface. In this case the Gauss curvature is $K = - \mu^2 < 0$, with constant $\mu$, and one can choose as parametric curves the asymptotic lines, for which $d_{11}=d_{22}=0$. Setting $\Delta^2= \textrm{Det}\left(g_{\mu\nu}\right)$, we see that
\beq
K=-\frac{d_{12}^2}{\Delta^2}\;.
\eeq
After some manipulations \cite{rogers2002backlund}, it can be shown that in this case the Mainardi-Codazzi equations imply
\beq
\Gamma_{12}^1 = \Gamma_{12}^2 = 0 \quad \Longrightarrow \quad \frac{\partial}{\partial z^2}\left(g_{11}\right) =\frac{\partial}{\partial z^1}\left( g_{22}\right) = 0 \;.
\eeq
Defining the angle $\omega$ between the parametric lines as
\beq
    \cos\omega = \frac{g_{12}}{\sqrt{g_{11}g_{22}}}\;,\qquad \sin\omega = \frac{\Delta}{\sqrt{g_{11}g_{22}}}\;,
\eeq
we have the following expression for the fundamental forms
\begin{align}
    \textsc I &= g_{11}\left(dz^1\right)^2 + 2\sqrt{g_{11}g_{22}}\cos\omega \,dz^1 dz^2 +g_{22}\left(dz^2\right)^2\;, \\
    \textsc{II} &= 2\mu\sqrt{g_{11}g_{22}}\sin\omega \,dz^1 dz^2\;.
\end{align}
Now, given the (anti-)holomorphicity of $g_{11}$ and $g_{22}$ we can rescale the variables $z^\mu$ to $z'^\mu=\sqrt{g_{\mu\mu}}z^{\mu}$ (no summation on repeated indices here) in terms of which one has\footnote{This corresponds to a parametrization of the surface by arc-length along the asymptotic lines.}
\begin{align}
    \textsc I &= \left(dz'^1\right)^2 - 2\cos\omega \,dz'^1 dz'^2 +\left(dz'^2\right)^2\;, \\
    \textsc{II} &= 2\mu \sin\omega \,dz'^1 dz'^2\;.
\end{align}
It is possible to show that the GMC system (\ref{eq:GMC}) reduces to the sine-Gordon equation
\begin{equation}
    \frac{\partial}{\partial z'^1}\frac{\partial}{\partial z'^2}\omega = \mu^2\sin\omega\;.
\end{equation}
Let us now consider the matrices $U_\mu$
\beqa
\label{eq:U1U2}
U_1 &=& \left(\begin{array}{c c c}
    \omega_1\,\cot\omega & -\omega_1\,\csc\omega & 0\\
    0 & 0 & \mu\sin\omega\\
    \mu\cot\omega & -\mu\csc\omega & 0
\end{array}\right)\;, \notag \\
 U_2 &=& \left(\begin{array}{c c c}
    0 & 0 & \mu\sin\omega\\
    -\omega_2\,\csc\omega & \omega_2\,\cot\omega & 0\\
    -\mu\csc\omega & \mu\cot\omega & 0
\end{array}\right)\;,
\eeqa
where $\omega_\mu = \frac{\partial}{\partial z^\mu}\omega$. The matrices (\ref{eq:U1U2}) do not belong to $\mathfrak{su}\left(2\right)$, as we would expect, and contain no trace of the spectral parameter $\lambda$. We can fix these apparent problems by the following considerations. First we notice that the triple $\sigma = \left\lbrace\boldsymbol{r}_1,\boldsymbol{r}_2,\boldsymbol{n}\right\rbrace$ is not orthonormal. However, the rotation
\beq
    \sigma \;\longrightarrow\; \tilde{\sigma} = M \sigma\;,\qquad M = \left(\begin{array}{c c c}
    1 & 0 & 0\\
    -\cot\omega & \csc\omega & 0\\
    0 & 0 & 1
\end{array}\right)\;,
\eeq
which corresponds to a gauge transformation on the matrices $U_\mu$
\beq
    U_\mu\;\longrightarrow\;\tilde U = \left(\partial_{\mu}M\right)M^{-1} + M U_{\mu}M^{-1}\;,
\eeq
leaves the compatibility equation -- the sine-Gordon equation -- invariant and maps (\ref{eq:U1U2}) into
\beq
\tilde U_1 = \left(\begin{array}{c c c}
    0 & -\omega_1 & 0\\
    \omega_1 & 0 & \mu\\
    0& -\mu & 0
\end{array}\right)\;, \qquad \tilde U_2 = \left(\begin{array}{c c c}
    0 & 0 & \mu\sin\omega\\
   0 & 0 & -\mu\cos\omega\\
    -\mu\sin\omega & \mu\cos\omega & 0
\end{array}\right)\;,
\eeq
which now belong to the $\mathfrak{su}\left(2\right)$ algebra. Finally, the spectral parameter can be recovered by noticing that the sine-Gordon equation is invariant under the following transformation
\beq
    \left(z'^1,z'^2,\mu\right)=  \left(\alpha \tilde{z}^1,\beta \tilde{z}^2,\frac{1}{\sqrt{\alpha\beta}}m\right)\;,
\eeq
for any constant $\alpha$ and $\beta$. Choosing $\alpha = \sqrt{2}m$ and $\beta = \sqrt{2}\frac{m}{\lambda^2}$ and writing $\omega = \beta \phi$, we obtain
\begin{align}
    \textsc I &= 2m^2\left(\left(d\tilde{z}^1\right)^2 - \frac{2}{\lambda^2}\cos\beta \phi \,d\tilde{z}^1 d\tilde{z}^2 + \frac{1}{\lambda^4}\left(d\tilde{z}^2\right)^2\right)\;, \\
    \textsc{II} &= 2\sqrt{2}\frac{m^2}{\lambda} \sin\beta\phi \,d\tilde{z}^1 d\tilde{z}^2\;,
\end{align}
which coincides with the quadratic forms (\ref{eq:I_sG}, \ref{eq:II_sG}).\\
Finally, as another interesting example of integrable model associated to embedded surfaces, let us briefly discuss a constant mean curvature surface, {\it i.e.} a surface such that $H=\textrm{const.}$. In this case one can choose conformal coordinates, in which the fundamental forms simplify to
\begin{align}
    \textsc I &= \frac{2}{H^2} e^{\omega} dz^1 dz^2 \;, \\
    \textsc{II} &= \frac{1}{H}\left[ A_1 \left(dz^1\right)^2 + 2e^{\omega} dz^1dz^2 + A_2 \left(dz^2\right)^2\right]\;.
\end{align}
Some simple computation shows that the GCM equations are equivalent to the system
\begin{align}
    \frac{\partial}{\partial z^1}\frac{\partial}{\partial z^2}\omega &= e^{\omega} - A_1 A_2 e^{-\omega}\;, \\
    \frac{\partial}{\partial z^2} A_1 &= \frac{\partial}{\partial z^1} A_2 = 0\;,
\end{align}
which is known as \emph{modified sinh-Gordon equation}. Its Gauss curvature is
\begin{equation}
    K = H^2\left(1-A_1 A_2 e^{-2\omega}\right) \;.
\end{equation}
Rescaling the field as $\omega \rightarrow \omega + 2\log H$, the functions $A_i$ as $A_i \rightarrow H A_i$ and sending $H\rightarrow 0$ yields a minimal surface and reduces the GMC system to Liouville equation
\begin{equation}
    \frac{\partial}{\partial z^1}\frac{\partial}{\partial z^2}\omega = K e^{\omega}\;,\quad K = -A_1 A_2 e^{-2\omega}\;.
\end{equation}
\section{Computation of the fundamental quadratic forms from sine-Gordon ZCR}

While in the preceding appendix we presented the derivation of soliton equations starting from the basic geometric data of some particular surface, here we wish to follow the reverse path and explicitly show how to obtain the forms (\ref{eq:I_sG}, \ref{eq:II_sG}) starting from sine-Gordon ZCR (\ref{eq:sG_ZCR_1}, \ref{eq:sG_ZCR_2}). First of all we need to find a basis of $\mathfrak{su}\left(2\right)$ with respect to the Killing form
\begin{equation}
    \left(a,b\right)_K = \textrm{Tr}\left(\textrm{\bf Ad}(a)\,\textrm{\bf Ad}(b)\right)\;,\quad a,b\in\mathfrak{su}\left(2\right)\;.
\end{equation}
In the adjoint representation one has $T^{i} = \textrm{\bf Ad}\(\cS^i\)$, with
\begin{equation}
    T^1 = \left(\begin{array}{c c c}
        0 & 1 & 0 \\
        -1 & 0 & 0 \\
        0 & 0 & 0
    \end{array}\right)\;,\quad T^2 = \left(\begin{array}{c c c}
        0 & 0 & -1 \\
        0 & 0 & 0 \\
        1 & 0 & 0
    \end{array}\right)\;,\quad T^3 = \left(\begin{array}{c c c}
        0 & 0 & 0 \\
        0 & 0 & 1 \\
        0 & -1 & 0
    \end{array}\right)\;,
\end{equation}
and
\begin{equation}
    \left(T^i,T^j\right)_K = -2\delta^{ij}\;.
\end{equation}
The orthonormal basis is easily found to be
\begin{equation}
    \mathbf e^i = \frac{\mathbbm{i}}{\sqrt{2}} \cS^i\;,
\end{equation}
and we see that for a pair of matrices $A$ and $B$ belonging to the $2$-dimensional representation of $\mathfrak{su}\left(2\right)$, one has
\begin{equation}
    \left(A,B\right)_{K} = 4 \; \textrm{Tr}\(AB\)\;.
\end{equation}
Now we need the partial derivatives of $r$ (\ref{eq:positionvectoring})
\begin{equation}
    r = \Phi^{-1}\frac{\partial}{\partial \lambda}\Phi\;\Longrightarrow \; \frac{\partial}{\partial z^{\mu}} r = \Phi^{-1}\frac{\partial}{\partial \lambda}\left(L_{\mu}\Phi\right) - \Phi^{-1}L_{\mu}\Phi\Phi^{-1}\frac{\partial}{\partial \lambda}\Phi\;,
\end{equation}
where we have used the linear system $\partial_\mu\Phi = L_{\mu} \Phi$. We have then
\begin{equation}
    \frac{\partial}{\partial z^{\mu}} r = \Phi^{-1}\frac{\partial L_{\mu}}{\partial \lambda} \Phi\;.
\end{equation}
We can immediately compute the metric tensor $g_{\mu\nu}$
\begin{equation}
    g_{\mu\nu} = \left(\frac{\partial r}{\partial z^{\mu}} , \frac{\partial r}{\partial z^{\nu}}\right)_K = 4\, \textrm{Tr}\(\frac{\partial r}{\partial z^{\mu}} \frac{\partial r}{\partial z^{\nu}}\) = 4\,\textrm{Tr}\(\frac{\partial L_{\mu}}{\partial \lambda} \frac{\partial L_{\nu}}{\partial \lambda}\)\;.
\end{equation}
Inserting the expressions (\ref{eq:sG_ZCR_1}, \ref{eq:sG_ZCR_2}) we obtain
\begin{equation}
    g_{\mu\nu} = 2m^2\left(\begin{array}{c c}
        1 & -\frac{1}{\lambda^2} \cos\left(\beta \phi\right) \\
        -\frac{1}{\lambda^2} \cos\left(\beta \phi\right) & \frac{1}{\lambda^4}
    \end{array}\right)_{\mu\nu}\;.
\end{equation}
The second derivatives of $r$ follow from simple computations
\begin{equation}
     \frac{\partial}{\partial z^{\mu}} \frac{\partial}{\partial z^{\nu}} r = \Phi^{-1}\left(\frac{\partial}{\partial z^{\nu}} \frac{\partial L_{\mu}}{\partial \lambda} + \left[\frac{\partial L_{\mu}}{\partial \lambda} , L_{\nu}\right]\right)\Phi\;.
\end{equation}
The matrix version of the unit normal is 
\begin{equation}
    n = \sum_{i=1}^{3} n_i \cS^i = \frac{1}{2\sqrt{2}}\frac{\left[\frac{\partial r}{\partial z^{1}},\frac{\partial r}{\partial z^{2}}\right]}{\sqrt{\textrm{Det}\left( \left[\frac{\partial r}{\partial z^{1}},\frac{\partial r}{\partial z^{2}}\right] \right)}}\;.
\end{equation}
We  obtain that
\begin{equation}
    \textrm{Det}\left( \left[\frac{\partial r}{\partial z^{1}},\frac{\partial r}{\partial z^{2}}\right] \right) = \left( \frac{m^2}{2\lambda^2}\sin\left(\beta \phi\right)\right)^2\;.
\end{equation}
We can finally compute the second fundamental tensor
\begin{align}
    d_{\mu\nu} &= \left(\frac{\partial}{\partial z^{\mu}} \frac{\partial}{\partial z^{\nu}} r , n\right)_K \notag \\
     &= \frac{1}{\sqrt{2}}\frac{\lambda^2}{m^2 \sin\left(\beta\phi\right)} \textrm{Tr}\left(\left[ \frac{\partial L_{1}}{\partial \lambda} , \frac{\partial L_{2}}{\partial \lambda} \right]\left(\frac{\partial}{\partial z^{\nu}} \frac{\partial L_{\mu}}{\partial \lambda} + \left[\frac{\partial L_{\mu}}{\partial \lambda} , L_{\nu}\right]\right)\right)\;.
\end{align}
The explicit expression is
\begin{equation}
    d_{\mu\nu} = \frac{\sqrt{2}m^2}{\lambda}\sin\left(\beta \phi\right) \left(\begin{array}{c c}
        0 & 1 \\ 1 & 0
    \end{array}\right)\;.
\end{equation}

\bibliography{Biblio3}

\end{document}